\documentclass[manuscript,screen]{acmart}

\AtBeginDocument{%
  \providecommand\BibTeX{{%
    \normalfont B\kern-0.5em{\scshape i\kern-0.25em b}\kern-0.8em\TeX}}}

\setcopyright{rightsretained}
\copyrightyear{2023}
\acmYear{2023}
\acmDOI{XXXXXXX.XXXXXXX}

\acmConference[Under Submission 'XX]{Make sure to enter the correct
  conference title from your rights confirmation emai}{2023}{}
\acmPrice{15.00}
\acmISBN{978-1-4503-XXXX-X/18/06}

\usepackage{ulem}
\usepackage{xcolor}
\usepackage{tabularx}
\usepackage{longtable}
\usepackage[caption=false]{subfig}
\usepackage{multirow}

\begin{document}

\title[The AI Ghostwriter Effect]{The AI Ghostwriter Effect: When Users Do Not Perceive Ownership of AI-Generated Text But Self-Declare as Authors}

\author{Fiona Draxler}
\email{fiona.draxler@ifi.lmu.de}
\orcid{0000-0002-3112-6015}
\author{Anna Werner}
\email{werner.anna@campus.lmu.de}
\orcid{0009-0009-0664-9074}
\affiliation{%
  \institution{LMU Munich}
  \city{Munich}
  \state{Bavaria}
  \country{Germany}
  \postcode{80337}
}

\author{Florian Lehmann}
\email{florian.lehmann@uni-bayreuth.de}
\orcid{0000-0003-0201-867X}
\affiliation{%
  \institution{University of Bayreuth}
  \city{Bayreuth}
  \state{Bavaria}
  \country{Germany}
  \postcode{95440}
}

\author{Matthias Hoppe}
\email{matthias.hoppe@ifi.lmu.de}
\orcid{0000-0002-5098-4824}
\author{Albrecht Schmidt}
\email{albrecht.schmidt@ifi.lmu.de}
\orcid{0000-0003-3890-1990}
\affiliation{%
  \institution{LMU Munich}
  \city{Munich}
  \state{Bavaria}
  \country{Germany}
  \postcode{80337}
}

\author{Daniel Buschek}
\email{daniel.buschek@uni-bayreuth.de}
\orcid{0000-0002-0013-715X}
\affiliation{%
  \institution{University of Bayreuth}
  \city{Bayreuth}
  \state{Bavaria}
  \country{Germany}
  \postcode{95440}
}

\author{Robin Welsch}
\email{robin.welsch@aalto.fi}
\orcid{0000-0002-7255-7890}
\affiliation{%
  \institution{Aalto University}
  \city{Espoo}
  \state{}
  \country{Finland}
  \postcode{FI-00076}
}

\renewcommand{\shortauthors}{Draxler et al.}

\def\subsectionautorefname{Section}
\def\subsubsectionautorefname{Section}
\def\sectionautorefname{Section}

\newcommand{\getting}{\textsc{Getting}}
\newcommand{\choosing}{\textsc{Choosing}}
\newcommand{\editing}{\textsc{Editing}}
\newcommand{\writing}{\textsc{Writing}}

\newcommand{\im}{\textsc{Interaction Method}}
\newcommand{\ims}{\textsc{Interaction Methods}}

\newcommand{\ai}{\textsc{AI}}
\newcommand{\human}{\textsc{Human}}

\newcommand{\aighostwriter}{\textit{AI Ghost\-writer Effect}}

\begin{abstract}
  Human-AI interaction in text production increases complexity in authorship. In two empirical studies (n1 = 30 \& n2 = 96), we investigate authorship and ownership in human-AI collaboration for personalized language generation. We show an \aighostwriter: Users do not consider themselves the owners and authors of AI-generated text but refrain from publicly declaring AI authorship. Personalization of AI-generated texts did not impact the \aighostwriter, and higher levels of participants' influence on texts increased their sense of ownership. Participants were more likely to attribute ownership to supposedly human ghostwriters than AI ghostwriters, resulting in a higher ownership-authorship discrepancy for human ghostwriters.
  Rationalizations for authorship in AI ghostwriters and human ghostwriters were similar. We discuss how our findings relate to psychological ownership and human-AI interaction to lay the foundations for adapting authorship frameworks and user interfaces in AI in text-generation tasks.
\end{abstract}



\maketitle

\section{Introduction}

Imagine your short visit to New York is coming to an end, and rather than spending time writing a postcard yourself, you might ask GPT-3 to write a personalized postcard for you. Would you sign this postcard with your own name? The use of personalized artificial intelligence (AI) in human-AI collaboration has the potential to significantly impact the concept of authorship. One example of such collaboration can be seen in using language generation models, such as GPT-3, to assist with writing tasks. Previous HCI research has investigated user perspectives on collaborative creative writing \cite{lee_coauthor_2022,ghajargar_redhead_2022,swanson_say_2012}, style transfer \cite{reif_recipe_2021}, text summarization \cite{goyal_news_2022}, and perceived authorship with AI suggestions \cite{lehmann_suggestion_2022}. However, it has not yet investigated how people attribute and declare authorship for the generated text. Previous research in the social sciences has investigated the prevalence and rationalization of ghostwriting \cite{gallicano2013ghost,carvalho_defying_2021,riley_crafting_1996}. Ghostwriting, as a practice of using text produced by someone without crediting, is common in some academic fields, such as medicine \cite{wislar_honorary_2011,gotzsche_ghost_2007}, but also in writing autobiographies or political speeches \cite{brandt_whos_2007}. Reasons for the use of ghostwriters include financial or political interests \cite{gotzsche_ghost_2007}, a lack of time and writing expertise \cite{brandt_whos_2007}, and the pressure to obtain gratification, e.g., good grades \cite{lines2016ghostwriters}.
Personalized Large Language Models (LLMs) are becoming accessible to the public and have the potential to become ubiquitous for everyday tasks that include text production. Therefore, it is necessary to understand how authorship is declared with personalized AI and to what extent AI is used similarly to human ghostwriters. This is particularly relevant when texts are personalized to users' experiences and individual writing styles, and the lines between user and AI contributions start to blur.

In this paper, we approach human-AI interaction for personalized LLMs from an authorship perspective. Thus, we are interested in investigating the psychological processes involved when attributing ownership and authorship for text generation with personalized AI, i.e., with models fine-tuned to an individual user. Note that this is different from the focus in ongoing legal and ethical discussions of AI-generated material that will yield a prescriptive framework, i.e., how authorship should be declared, and that focuses much on non-personalized material in academic and professional contexts \cite{samuelson_ai_2020}.
Consequently, we study authorship in a \textit{personal} writing context, where declared authorship reflects individual decisions rather than explicit guidelines and where personalization to individual writing styles is an emerging practice.

To tackle the topic of ownership and authorship in personalized LLMs, we conducted two empirical studies that show the \aighostwriter{} of personalized AI. We define this effect as the use of personalized generative AI without credit to the AI. Thus, the AI acts as a ghostwriter.
Overall, we found that users do not judge themselves to be the authors of AI-generated text but often refrain from publicly declaring authorship of AI. In Study 1, we compared whether personalization and different interaction methods affect perceived ownership and declared authorship. In Study 2, we replicate the \aighostwriter{} in a large sample and compare it to a human ghostwriter.

We show that personalization is not critical to the \aighostwriter{} (Study 1). Subjective control over the interaction and the content increases the sense of ownership. Moreover, people use similar rationalizations for AI ghostwriters and human ghostwriters (Study 2, pre-registered at \url{https://aspredicted.org/RKV_ZXX})\footnote{Note that both studies were conducted before AI crediting policies as put forward by Nature and arXiv, cf. \autoref{sec:ai_authorship_declaration}.}. From this, we motivate how to expand common authorship declaration frameworks (e.g., the CRediT taxonomy\footnote{\url{https://credit.niso.org}}, \cite{allen_how_2019,allen2014publishing,brand2015beyond}) to account for the support of AI from a user-centered perspective. Understanding the interplay between control, ownership, and authorship also informs the interaction design of future AI-supported text-generation systems.

\section{Related Work}

This section outlines the current landscape of text-generation models and working with text-generation systems from a technological and human-centered perspective. In addition, it discusses the concepts of ownership and authorship and how it is currently changing with automated generation.

\subsection{Generative Models and Personalization}

In recent years, various generative models have emerged for tasks such as text-to-text \cite{brown_language_2020} and text-to-image\footnote{\url{https://openai.com/dall-e-2/}} generation, music composition based on prompts\footnote{\url{https://mubert.com}}, and text simplification \cite{yang_towards_2021}. %
Other models generate images \cite{ruiz_dreambooth_2022}, poetry \cite{tikhonov2018guess}, rap lyrics \cite{potash2015ghostwriter}, or music \cite{huang_music_2018} in the style of previous works.
Generative models often implement a Transformer architecture that includes an encoder and a decoder, utilizing an attention mechanism that allows back-references to prior content \cite{vaswani_attention_2017}.
These models are typically trained to predict probable next tokens from the previous text. Thus, they learn connections between subsequent words and are well-suited for text generation.

In this work, we focus on text generation with GPT-3.
Released in 2020, GPT-3 currently is a particularly large and powerful language model \cite{dale_gpt3_2021,dehouche_plagiarism_2021}. It uses deep learning methods such as attention mechanisms and the Transformer architecture with autoregressive pretraining  \cite{brown_language_2020}. This architecture has improved the generation of long coherent text \cite{uchendu_turingbench_2021} because of the model's ability to give special attention to key textual features and to connect interrelated words over longer passages \cite{elkins_can_2020}. GPT-3 produces text\footnote{Primarily in English} that statistically fits well with a given natural language prompt, using its syntactic ability to associate words without understanding the semantics and context of the query \cite{floridi_gpt3_2020}. The text quality is often comparable to human-written texts in terms of grammar, coherency, and fluency, but it can also produce nonsensical or incorrect content \cite{uchendu_turingbench_2021, elkins_can_2020}. Like many other machine learning systems, GPT-3 may also reproduce biases found in its training data, such as racial or gender biases \cite{lucy_gender_2021,brown_language_2020}.
One use case in which GPT-3 performs well is ``few-shot learning,'' where demonstrations of input-output pairs for the task are included with the prompt \cite{brown_language_2020}. The choice of these in-context examples has a significant influence on the content and the quality of the generated text \cite{liu_what_2021}.

The same idea can also be used for personalization: the output of the system is conditioned by the information available about the user \cite{yang_towards_2021} and their context \cite{dudy_refocusing_2021} to better match generated texts to a writer's expectations and needs.
According to \citet{wang_automatic_2018}, personalization can happen at two levels: it can include factual knowledge such as personal data, user attributes, and user preferences. The second category is defined as stylistic modifiers, which can either be situational or personal. For example, \textit{Gmail Smart Compose} interpolates between a global and a personal model trained on individual sent messages \cite{chen_gmail_2019}.
Research in related AI-supported scenarios shows that personalization and adaptation can improve the perceived quality of results \cite{kim_designing_2019,peng_trip_2018} and engagement \cite{huang_effects_2023}. It is likely that similar effects also hold for text generation. However, it is challenging to match generated texts with a writer's values or perspective, which influences acceptance \cite{biermann_tool_2022}. Overall, LLMs can be considered a powerful and versatile technology for writing that could be improved with personalization.

\subsection{Interaction with Text-Generation Systems}

In HCI research, text generation has also been investigated and applied in user interfaces for text entry.
Originally, such ``intelligent'' or ``predictive'' text entry methods were developed as augmentative and alternative communication (AAC) tools to reduce manual efforts for people with cognitive or motor impairments. Concretely, this included suggesting whole words, which users can then select instead of typing them letter by letter~\cite{higginbotham_evaluation_1992, Fowler2015}. 
Later, similar approaches were applied towards the grand goal of efficiency~\cite{Kristensson2014}: The bottleneck of an idealized text entry method is the cognitive process of coming up with ideas, not the physical process of entering them. In systems such as Google's \textit{Smart Reply}, short text suggestions are designed to possibly skip manual writing altogether~\cite{Kannan2016smartreply}.

With the rise of large language models, we now see systems increasingly being conceptualized, presented and/or reflected on as \textit{writing assistants}, which may provide some degree of inspiration~\cite{lee_coauthor_2022, Singh2022elephant, Yuan2022wordcraft, Bhat2022,chung_intersection_2021}. 
Generative models could also enhance the versatility of interactive storytelling systems as proposed by \citet{swanson_say_2012}.
The UIs of such systems now explore designs beyond integration at the user's cursor in the text box or a ``suggestion bar'' above the keyboard (cf. \cite{Arnold2016phrases_vs_words, Bi2014, Buschek2021emails, Quinn2016chi}): For example, generated inspirational content is controlled and presented via sidebars~\cite{Yuan2022wordcraft}, visual story arcs~\cite{Chung2022talebrush}, manipulable metaphors~\cite{Gero2019metaphor}, and modalities beyond text (e.g. images, sound)~\cite{Singh2022elephant}.
In sum, the broader framing of writing assistance links traditional HCI questions (e.g. UI design factors) to emerging questions of authorship (cf.~\cite{lehmann_suggestion_2022}).

Indeed, many researchers (implicitly) touch upon such questions in their interaction studies, without addressing them explicitly in the study design, since that is not their focus:
For instance, \citet{lee_coauthor_2022} call their system ``CoAuthor'' and introduce measures of ``mutuality'' and ``equality'' of the text contributions by users and the AI. Similarly, \textit{WordCraft} is presented as a ``collaboration'' and ``conversation'' with the assisting language model system~\cite{Yuan2022wordcraft}. Focusing on input behavior with such systems, \citet{Buschek2021emails} quantitatively identified nine patterns, from ignoring suggestions to chaining several suggested phrases in a row. \citet{Arnold2016phrases_vs_words} indicated that phrases are perceived more as ideas, in contrast to words, which are seen more as predicted continuations. Related, \citet{Singh2022elephant} describes how writers actively invest into taking `leaps' to fit their story around a suggested phrase. Similarly, \citet{Dang2022uist} found evidence of writers changing their original text to influence AI-generated summaries, as well as interpreting these summaries as an external perspective on their text. Qualitatively, \citet{Bhat2022} dissect such AI influences on writers and their text through the lens of the cognitive process model of writing from \citet{Hayes2012}, which distinguishes between proposing ideas, translating those into text, and finally transcribing these with a given input method. Suggestions may influence all three of these processes. Recent work shows potential influences on the writer's opinion in this context~\cite{Jakesch2023opinionatedLLMs}. 

A specific writing use case that already benefits from ample AI support in practice is communication (e.g. chat messages, email). Research on AI support in this context coined the term ``AI-mediated communication'' (AI-MC)~\cite{Hancock2020JCMC}. Our postcard writing task represents a similar context. However, in contrast to some work on AI-MC, we do not examine the impact of AI on the receiver of the message or the sender-receiver relationship (cf.~\cite{Hohenstein2020moralcrumple, Jakesch2019aimediated, Mieczkowski2021cscw}). Instead, we study the sender's (self-)perception (of text ownership). In this regard, our study is related to AI-MC work by \citet{mieczkowski_examining_2022preprint}, who studied the impact of AI on the writers' perceived agency. While their think-aloud study of agency focuses on the process of writing with AI, we study a related ``final'' perception after creating the message -- i.e. perceived ownership.

In summary, research on text entry at the intersection of HCI and Natural Language Processing currently sees a strong shift toward much more extensive AI support for human writing. While many studies take note of aspects of (perceived) authorship along the way, a dedicated investigation that considers how authorship is attributed is still missing. This motivates our fine-grained analysis with multiple psychologically relevant variables (e.g., sense of ownership) and the comparison of different interaction designs in Study 1. 

\subsection{Declared Authorship, Copyright, and Ghostwriting in Response to Generative AI}
\label{sec:ai_authorship_declaration}

The concepts of authorship and copyright contrast the practice of ghostwriting. While intuitively, any creator of a text or artwork should also be declared its author and hold the copyright, this is not always the case. Ghostwriting is the practice of not crediting a content creator in the publication \cite{wislar_honorary_2011,claxton_scientific_2005}. Reasons for ghostwriting are a lack of time, background details, or rhetoric skills \cite{brandt_whos_2007,riley_crafting_1996}, situations in which a public persona's name as the author is essential for marketing \cite{carvalho_defying_2021,gallicano2013ghost}, peer pressure to boost reputation or credibility in science \cite{nylenna_authorship_2014, gotzsche2009should}, or financial gain \cite{gotzsche_ghost_2007,plos2009ghostwriting}. Unethical authorship practices such as ghostwriting have led to the development of authorship criteria \cite{claxton_scientific_2005}. For example, the International Committee of Medical Journal Editors (ICMJE) defined that someone should be declared an author of a work if they \textit{made a significant contribution to the conception of the work or analyzing data for the work}, \textit{drafted or critically revised the work}, \textit{approved the version to be published}, and \textit{are taking responsibility for all aspects of the work}\footnote{\url{https://www.icmje.org/recommendations/browse/roles-and-responsibilities/defining-the-role-of-authors-and-contributors.html}}. Other organizations, such as the ACM\footnote{\url{https://www.acm.org/publications/policies/roles-and-responsibilities\#h-criteria-for-authorship}}, define similar criteria.
In a slightly different approach, the CRediT taxonomy classifies different types of contributions to clearly state the role of each individual contributor \cite{allen_how_2019} and concepts as proposed by \citet{bd_solving_2016} allow interactive role declaration.
This is particularly important when declarations of authorship become complex, for example, when several parties are involved or when work is derived from previous work \cite{diakopoulos_evolution_2007}.
Nonetheless, detailed attribution is sometimes not even supported by deployment platforms \cite{cho_individuals_2022}.

Adding personalized generative AI models to this mix further complicates authorship declarations. On the one side, copyright and intellectual property become a debated legal issue \cite{pihlajarinne_aigenerated_2019,sturm_artificial_2019,franceschelli2022copyright} where computer algorithms or their developers but also creators of the training data may hold rights to the content \cite{asay_independent_2020,ihalainen_computer_2018,samuelson_ai_2020,franceschelli2022copyright}. Rights allocation may indeed be entirely subjective, as evidenced by strikingly different legislation across countries \cite{samuelson_ai_2020}. Developers of generative AI, on the other side, assign all rights to their users. OpenAI states that their customers using the text-generation model GPT-3 or the image-generation model DALL·E possess full rights over the produced content\footnote{\url{https://openai.com/api/policies/sharing-publication/\#content-co-authored-with-the-openai-api-policy}}, including commercial use. According to their guidelines, texts generated by GPT-3 must be attributed to the user's name or their company. It must be stated that the content is AI-generated, but OpenAI retains all rights, titles, and interests in the language model itself.

There are some instances in the practice of publication that can give hints on the perspectives of different stakeholders for authorship in generative AI. In journalism, AI contributions are sometimes but not always declared \cite{kent_ethical_2019,graefe_guide_2016,montal_i_2017}. For example, in his \textit{Guide to Automated Journalism}, \citet{graefe_guide_2016} notes that AI authorship declarations are important for transparency. \citet{montal_i_2017} suggest using bylines of automatically generated articles to reference software vendors and a mention in a disclosure statement for algorithmic text parts. In rare cases, AI contributors have also been listed as authors of a scientific publication \cite{writer2019lithium,kung_performance_2022,oconnor_open_2023}. This has motivated several scientific associations to publish policies that discourage or prohibit listing text generation tools as an author, e.g., Nature\footnote{\url{https://www.nature.com/nature/for-authors/initial-submission}} and arXiv\footnote{\url{https://info.arxiv.org/help/moderation/index.html\#policy-for-authors-use-of-generative-ai-language-tools}}; they recommend adding the tools to a Methods section. Note that these do not distinguish between text-generating AI and personalized AI. Therefore, guidelines for the application of generative AI for text generation are largely applied to the professional context and are not nuanced enough to consider personalized generative AI.

While normative frameworks for specific application contexts are developing as a reaction to advances of LLMs, dedicated frameworks for personalized AI text generation are still missing. In particular, it is unclear how the use of text-generating AI is distinct from ghostwriting. Studying authorship declaration for personalized text-generating AI in users and comparing this to common ghostwriting practices can contribute to this debate by providing empirical data.

\subsection{AI Support and Sense of Ownership}
\label{sec:ownership}

Authorship is not only a question of declaration but also of individual judgment.
Whether someone feels like an author depends on factors such as their perceived control and agency.
Feelings of ownership---of objects, but also intangible entities such as ideas---can develop through creation and control (\textit{psychological ownership}; \cite{kim_appropriate_2016,pierce_state_2003}).
For example, in a video remixing context, \citet{diakopoulos_evolution_2007} found that creators often felt they had to make substantial changes to previous work before they could see it as their own. This indicates again that one's investment is essential for perceived ownership \cite{pierce_state_2003}.
Research on perceived authorship with generative AI is still scarce. Notably, \citet{lehmann_suggestion_2022} could show a correlation of authorship with control in interaction design: In their study, perceived authorship was stronger when participants wrote their own texts than when they received text suggestions.
In an interview study published in 2017, members of news organizations did not perceive algorithms as authors \cite{montal_i_2017}. Instead, they mentioned the respective journalistic organization or people who contributed to creating the algorithm, suggesting an anthropomorphic image of authorship. Therefore, the writing process and the resulting product must be regarded separately. In our paper, we, thus, address the interplay between the two dimensions in terms of perceived ownership and declaration of authorship.

\subsection{Summary of Related Work}
There is an ongoing debate about the attribution of authorship in the context of text-generating AI that mirrors ghostwriting practices. The debate is even more complex for personalized generative AI and for non-professional domains. %
HCI can contribute to this discourse by investigating how the design of the interaction with AI affects authorship declaration and related psychological phenomena, the most important being ownership. Empirical data is needed to ground the debate from an empirical user-based perspective and provide guidelines for designers of interfaces to text-generating AI.

\section{Research Questions and Hypotheses}
\label{sec:research_questions}

We pose the following research questions: 

\begin{enumerate}
    \item [\textbf{RQ1}:] Does the sense of ownership match the declaration of authorship for personalized AI-generated texts?
    \item [\textbf{RQ2}:] Does the level of influence in human-AI interaction affect the sense of ownership?
    \item [\textbf{RQ3}:] Does the sense of ownership depend on the quality of personalization?
    \item [\textbf{RQ4}:] In what ways does the \aighostwriter{} differ from human ghostwriters?
\end{enumerate}

Our paper aims to investigate what affects authorship attribution for texts generated with personalized LLMs from an HCI perspective. First, we identify the psychological processes involved when attributing authorship in human-AI interaction. Note that we consider the process of attributing authorship (from the \textit{sense of ownership} to the \textit{declaration of authorship}), the perceived degree of control in human-AI interaction, the subjective quality of personalization, and the feeling of leading the human-AI interaction.
In other words, the sense of ownership addresses the subjective side: does a person feel that they are the author of a text and that they own it? This also encompasses a sense of control over the product, as the two concepts are tightly linked (cf. \autoref{sec:ownership}). The declaration of authorship, on the other hand, refers to the entity a text is attributed to, e.g., in the header or byline.

If the AI system, here GPT-3, is used much like a human ghostwriter, then we hypothesize that there is an \aighostwriter{}:

 \begin{enumerate}
    \item[\textbf{H1.1:}]
     participants will consider the AI (and not themselves) to be the owner of personalized AI-generated texts but
    \item[\textbf{H1.2:}] they will not declare AI authorship or AI support when publishing personalized AI-generated texts.
 \end{enumerate}

Second, we want to investigate how common interaction methods affect the process of attributing authorship. This is important as designers of interfaces are faced with a large body of normative statements on authorship with little knowledge of how user interfaces affect the declaration of authorship and what user experience they create during human-AI interaction.

To effectively design interactive LLM-based applications with authorship in mind, we have to know how different interaction methods affect authorship for personalized AI-generated texts. \citet{lehmann_suggestion_2022} found that the sense of authorship positively correlates with the degree of influence over the AI contribution. Here, objective control refers to the degree of influence that users have over the AI-generated text, e.g., by employing interaction methods categorized as \writing{}, \editing{}, \choosing{}, and \getting, while perceived control refers to the subjective level of being able to affect the generated text a priori. Leadership, in turn, refers to the user's perceived initiative to shape the outcome, i.e., agency during the interaction.

Thus, we also assess perceived control and leadership, and we hypothesize that

 \begin{enumerate}
    \item[\textbf{H2.1:}]
    the degree of influence over the generated text, as realized by different interaction methods, affects the sense of control, and 
    \item[\textbf{H2.2:}]
    the degree of influence over the generated text affects the sense of leadership. 
 \end{enumerate}

Third, we investigate how AI personalization affects this process. Here, personalization refers to the customization of the AI model using fine-tuning to suit an individual user's specific preferences and writing style. 
We explicitly investigate the case of \textit{personalized} text generation because, in this case, the lines between the AI model and the user blur. However, recent studies have found that usability and user experience do not depend on the quality of personalization and adaptation for fine-tuned AI systems \cite{vaccaro_illusion_2018}. Even non-adaptive systems are deemed adaptive and personalized if introduced as such \cite{kosch_placebo_2022}. If this holds for personalized LLMs, merely labeling the AI as personalized (i.e., \textit{placebo-personalization}) should yield comparable effects on the sense of ownership.

We consider two levels of personalization quality -- personalization by fine-tuning and placebo-personalization -- and we hypothesize that
\begin{enumerate}
    \item[\textbf{H3:}]
     the sense of ownership is independent of the quality of personalization for the AI model.
\end{enumerate}

Fourth, we explore how using AI is qualitatively similar or distinct from the common practice of employing human ghostwriters. While ghostwriting has been a common practice in various fields like literature, speeches, and even academia, the advent of AI changes the dynamics of this process as generative AI models are easily accessible by a broad base of users.

Studies have shown that people are more inclined to exploit algorithms or AI models than humans \cite{karpus_algorithm_2021}. This phenomenon has implications for authorship attribution, the sense of ownership, and how these AI-assisted texts are presented to the public.  We posit the following hypotheses:
 
 \begin{enumerate}
    \item[\textbf{H4:}] The sense of ownership is lower for texts written by a human ghostwriter than for texts written by an AI ghostwriter.
      \item[\textbf{H5:}] People are more likely to attribute (co-)authorship to a human ghostwriter than to an AI ghostwriter.
    \item[\textbf{H6:}] When using AI-generated texts, people do not always declare the AI as a (co-)author, even when they do not have a sense of ownership (\aighostwriter{}). This serves to replicate H1, but without considering personalization.
  
\end{enumerate}

We test H1 - H3 in Study 1. We varied the level of influence over the AI-generated texts in the task of writing and publicly uploading a postcard from New York by employing different \ims{} (within-subject factor). Based on the correlation of influence and sense of ownership suggested in \cite{lehmann_suggestion_2022,pierce_state_2003}, we defined four methods in descending order of the level of user influence: \writing{}, \editing{}, \choosing{}, and \getting{}. We generated texts with personalization or placebo-personalization (fine-tuned for individual participants vs. base model; between-subject factor). %
The postcard writing scenario represents a personal writing task that people are generally familiar with. Postcards are individually written but often describe similar experiences, e.g., famous sights or local food. This enables us to apply placebo personalization. Posting postcards on a blog-style website was chosen to diminish the effect of social expectations a potential postcard recipient might have regarding authorship declaration.

To answer RQ4 and to replicate and extend Study 1, we conducted Study 2 (pre-registered, see \url{https://aspredicted.org/RKV_ZXX}\footnote{Note that we have slightly adapted the wording to match the manuscript.}), comparing AI-supported writing to the case of a human author supporting the writing task. 

In summary, Study 1 aims to examine the influence of control, personalization, and the sense of ownership in the context of AI text generation. Its purpose is to introduce the \aighostwriter{}, which refers to the pattern of users not attributing authorship or declaring any form of AI support for personalized AI-generated text.
The results from this study inform Study 2, which delves deeper into the dynamics between AI and human ghostwriting, further expanding our understanding of authorship attribution practices in human-AI interaction in comparison to human-human interaction for writing. Thus, Study 2 replicates Study 1 and sheds light on the quantitative differences in declaring authorship for AI and human ghostwriters.

\section{Study 1: Sense of Ownership versus Declared Authorship}
We varied the level of influence over the AI-generated texts in the postcard writing task by employing different \ims{}. In Study 1, our primary goal was to evaluate how participants attribute ownership to personalized AI-generated texts, whether a sense of control in the interaction affects that, and how participants declare authorship for different \ims{}. The study was conducted in July 2022.

\subsection{Method}
We implemented four \ims{} that vary with regard to user influence over the output from \writing{} a text manually (full influence) to \getting{} a text that was fully written by the AI (no influence). First, a sample of 30 participants filled out a survey that enabled us to fine-tune the parameters of the text-generation model so the generated texts match individual participants' writing styles. Then, the same participants interacted with an actually personalized or a placebo-personalized model with all four \ims{}, evaluated them, and were instructed to publish the texts online.

After approximately two days, we explored whether participants could still differentiate texts generated with the different \ims{}. This was added to evaluate the self-relevance of the generated texts. If participants could consistently differentiate between their own and other texts, this may indicate noticeable differences in their relation to the text as a function of the interaction method. Note that for the sake of brevity and because a discussion on the self-relevance of personalized AI-generated text would diverge the paper's focus, we report the results of this exploratory part of Study 1 as supplementary material. 

Therefore, the study had a 4 (\im) $\times$ 2 (personalization vs. placebo-personalization) design. In the personalization condition, GPT-3 was fine-tuned to the user's input; in the placebo-personalization, no fine-tuning was applied on the technical side. In both groups, participants were told they were using a personalized AI model.

\subsubsection{Implementation of the Interaction Methods}
\label{sec:system_design}

\begin{figure*}
    \centering
    \subfloat[\writing]{
        \includegraphics[width=.42\textwidth]{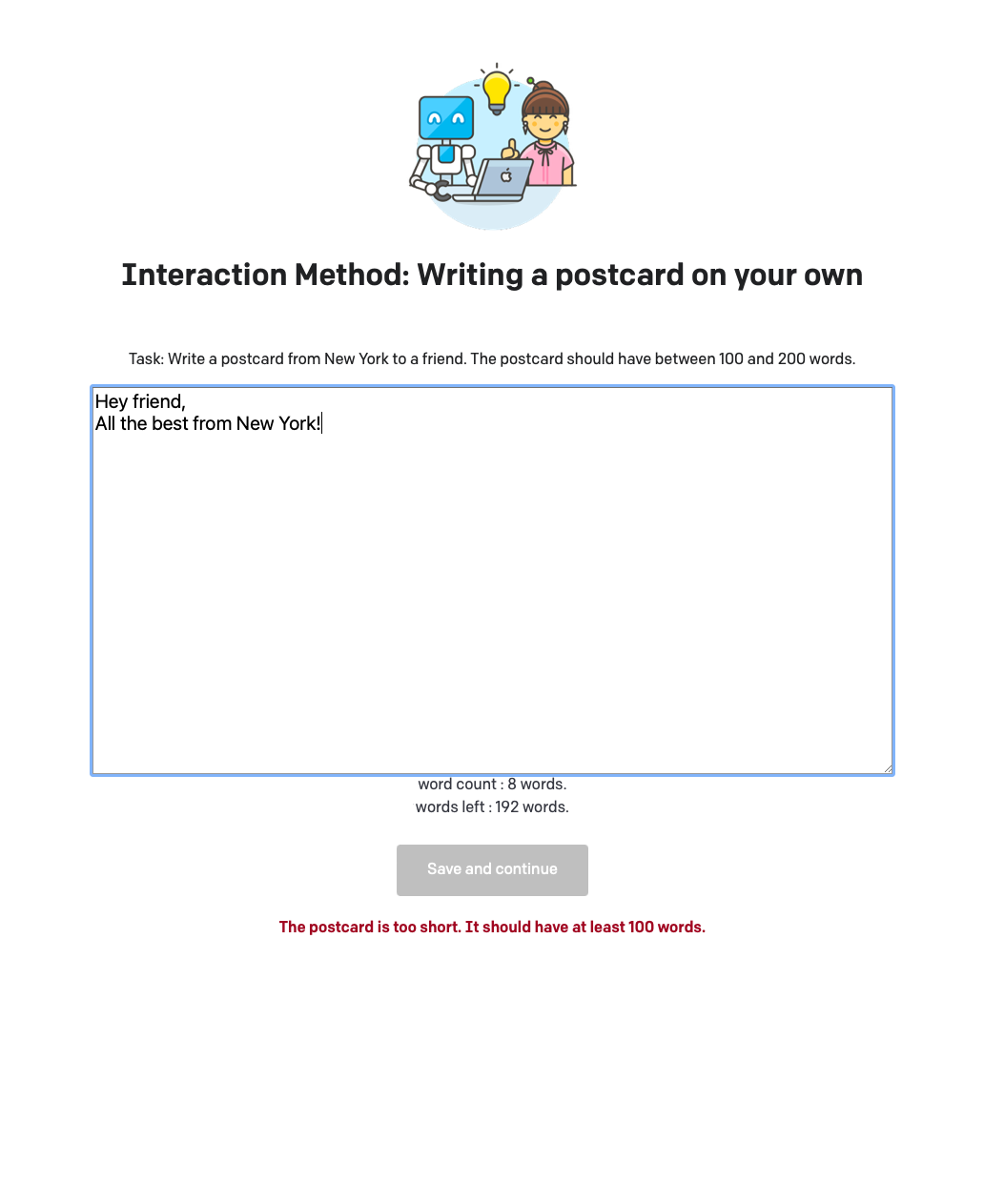}}\label{fig:im_writing}\qquad
    \subfloat[\editing]{
        \includegraphics[width=.42\textwidth]{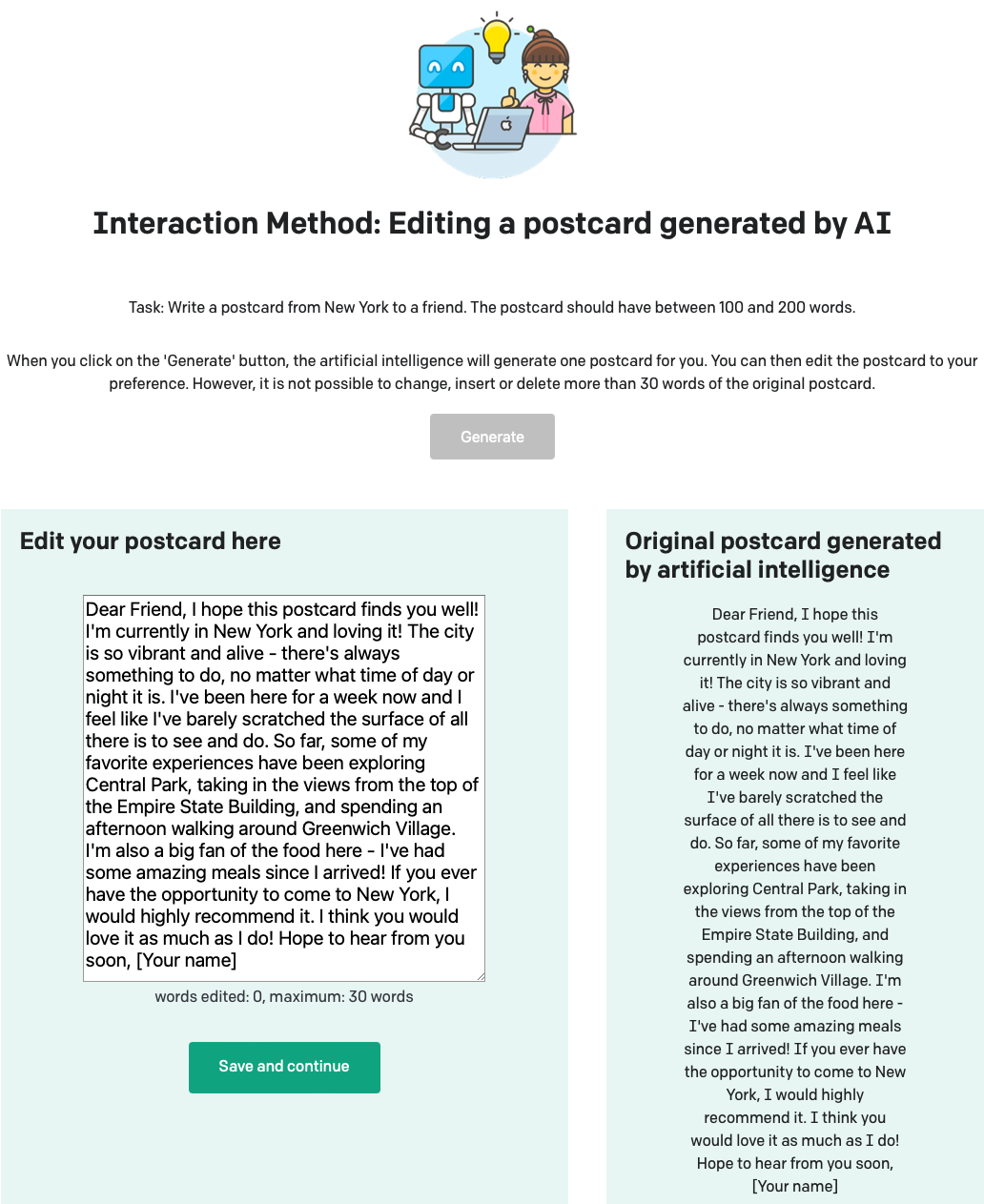}}\label{fig:im_editing}
    
    \subfloat[\choosing]{
    \includegraphics[width=.42\textwidth]{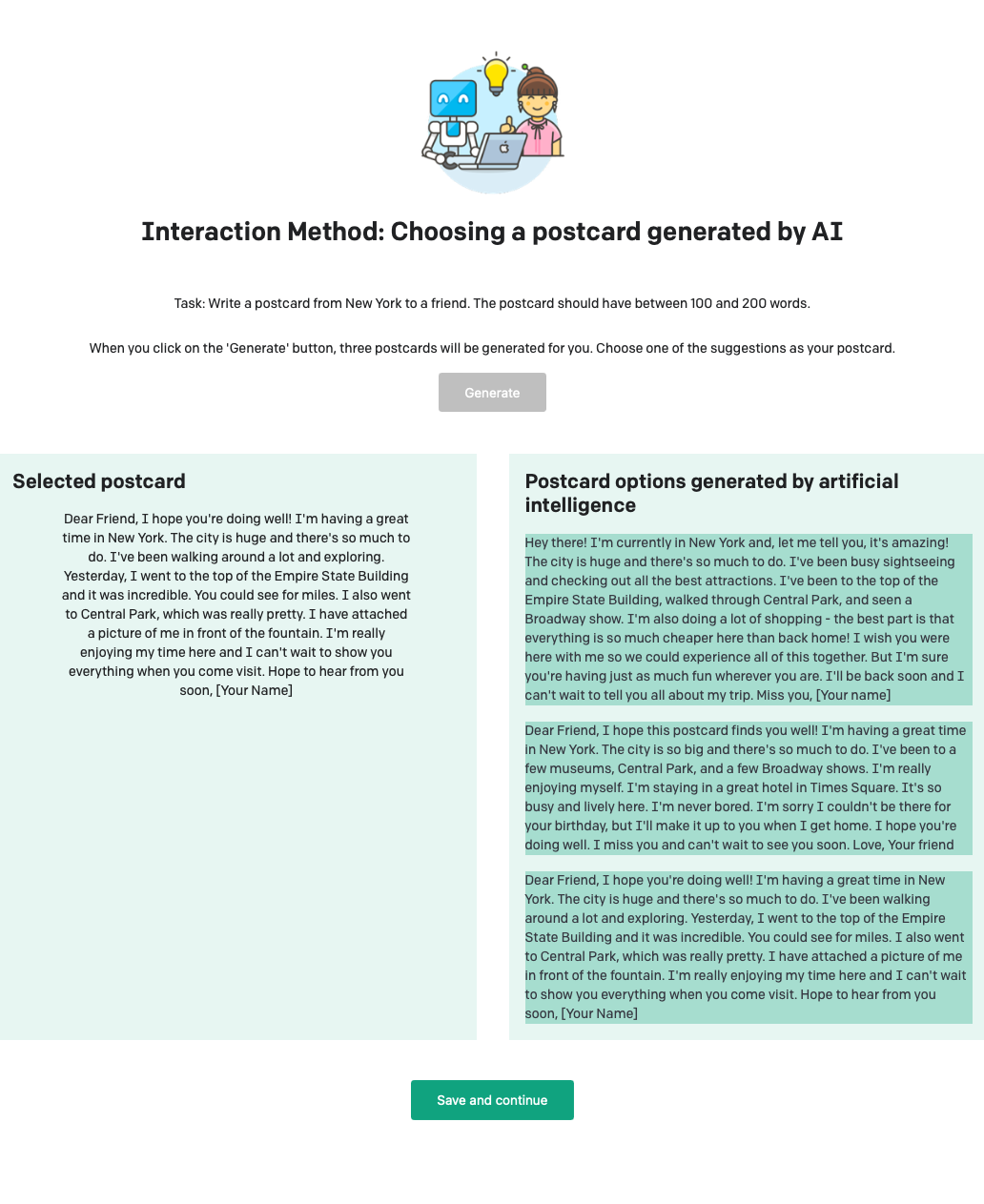}}\label{fig:im_choosing}\qquad
    \subfloat[\getting]{
        \includegraphics[width=.42\textwidth]{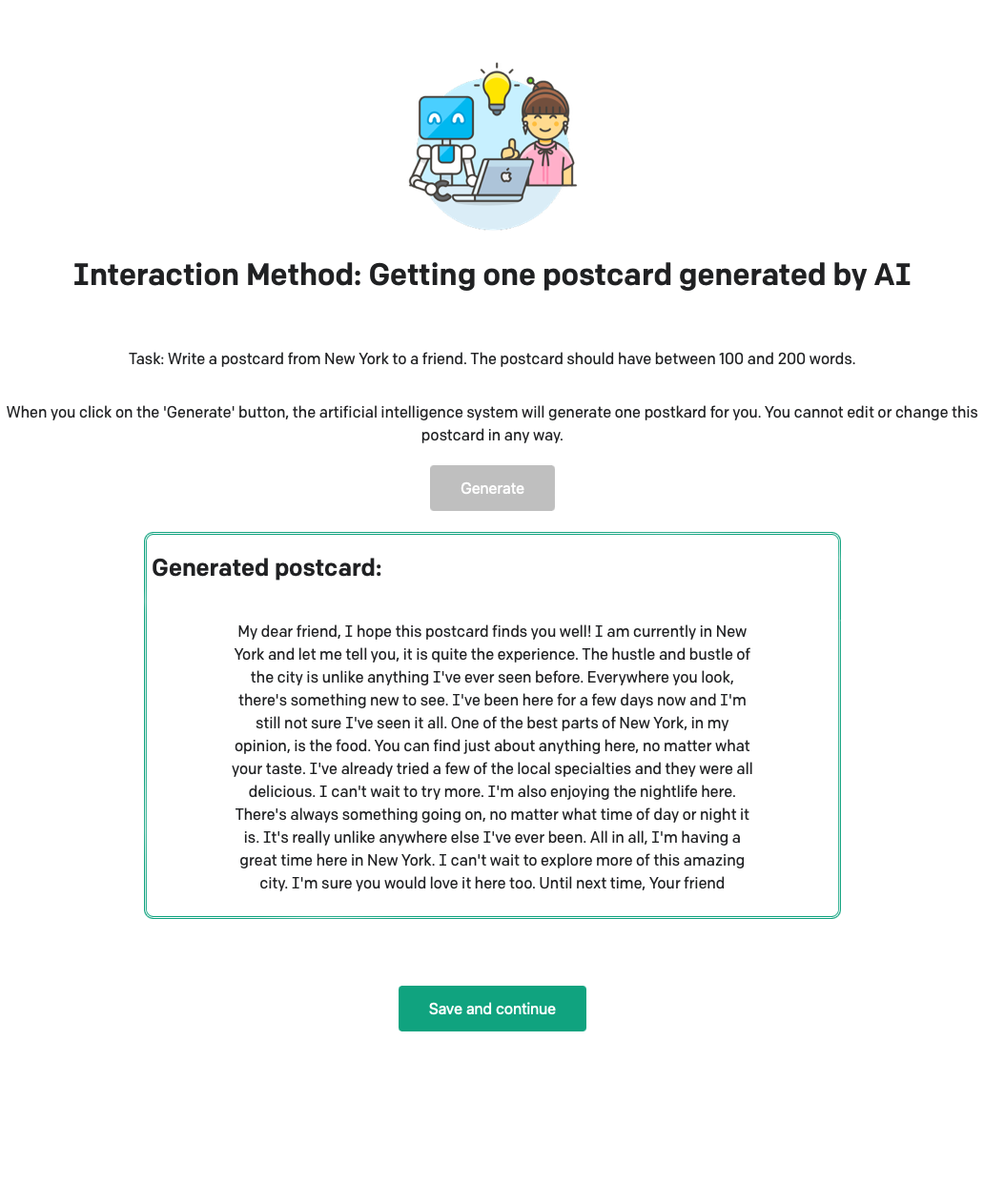}}\label{fig:im_getting}
    \caption{Screenshots of the \ims{} in Study 1. Human-robot-collaboration icon licensed from Iconfinder \cite{webalys2023working}.}
    \label{fig:im_screenshots}
\end{figure*}

We implemented the four \ims{} (IMs) in a web-based text-generation platform. Each IM represented a different level of the participants' influence over the text. See \autoref{fig:im_screenshots} for screenshots of the web prototypes. We describe the four IMs in more detail below.
The levels of control are derived from \cite{edwards_transparency_2020} and \cite{lee_coauthor_2022}.

\paragraph{(1) \writing{} a postcard.}
In this baseline condition, participants wrote a postcard manually without AI assistance, i.e., they fully influenced the results. They were required to enter between 100 and 200 words. This is equivalent to \textit{Level 0: Fully human-written language} in \cite{edwards_transparency_2020}.

\paragraph{(2) \editing{} a postcard.}
Participants were provided with an AI-generated postcard but could change (insert, delete, or replace) up to 30 words. With this method, participants could manually influence the text. The AI provided a first draft of the postcard. The specific editing limits were adjusted based on a pretest. This method represents a typical AI-supported scenario as in the \textit{Edit} phase in \cite{lee_coauthor_2022}.

\paragraph{(3) \choosing{} a postcard.}
Participants were presented a set of three AI-generated postcards and were asked to choose one. Participants had no manual influence on the text with this method. However, they could still choose the postcard version they considered most appropriate. This corresponds to \textit{Level 3: Conditional automation} in \cite{edwards_transparency_2020} and is similar to the \textit{Get Suggestions} phase in \cite{lee_coauthor_2022}.%

\paragraph{(4) \getting{} a postcard.}
Participants were presented one AI-generated postcard only. In contrast to the other methods, participants had no influence over the postcard text. Conversely, this represents the highest level of AI influence as the generator of the final postcard. This method implements \textit{Level 4: High automation}, i.e., generation without human supervision \cite{edwards_transparency_2020}.\medskip

The texts were generated prior to being shown in the study. Nevertheless, they were presented 3 seconds after the participants had clicked on a ``generate'' button to create the illusion that the postcard was actually generated on demand.

\subsubsection{Procedure} %
\label{sec:procedure_s1}

The study consisted of three main parts: (1) a fine-tuning survey for personalization, followed by background processing for generating personalized texts, (2) postcard text production with the interaction methods (\autoref{sec:system_design}), followed by an authorship and interaction questionnaire (see \autoref{tab:study1_measures}, Part 2), and (3) a final questionnaire on text recognition, text rating, and general conclusions. This final questionnaire was deployed two days later. %
In accordance with the declaration of Helsinki, we informed participants about their rights and the study procedure and only started the study if they gave their consent to take part in it.

\paragraph{Part 1: Fine-Tuning Survey and Background Processing}
For collecting fine-tuning input, we first asked participants to write responses of approximately 330-800 characters to 18 different prompts. The style and theme of the prompts resembled the prompt that we used for postcard generation in Part 2 (see below). Examples of prompts are ``Please describe yourself in about 5 sentences.'' and ``What was your favorite vacation?'' %
The participants were then asked to upload five texts they had previously written (e.g., blog posts or emails to friends) and match each one with a suitable prompt.
The full list of prompts is shown in \autoref{tab:study1_measures}.

Once the fine-tuning questionnaire was complete, we performed the background processing.
We randomly assigned half the participants to the personalized condition and the other half to the placebo-personalized condition. %
We personalized Davinci models\footnote{\url{https://beta.openai.com/docs/guides/fine-tuning}} with the data of the participants in the personalization group and generated five individual postcards for the postcard prompt (``Write a postcard from New York to a friend. The postcard should have between 100 and 200 words.''; three postcards for \choosing{} and one each for the other \ims{}).
For the placebo-personalization group, we randomly alternated between three distinct text sets that contained 5 of the 15 pre-generated texts\footnote{Technical details and the text sets are provided as supplementary material.}.

All participants responded to the fine-tuning questions regardless of their assigned personalization group, and they all had to wait for their invitation to Part 2. This served to mitigate response bias and the placebo effect in human-computer interaction \cite{kosch_placebo_2022,Caraban2019}.

\paragraph{Part 2: Writing \& Uploading, Authorship Evaluation}
One day later, participants started Part 2 of the study.
Here, their task was to create four postcards addressed to a friend as if they were currently visiting New York and upload the created postcards to a blog-style website. They pasted the postcard texts into a text field and provided metadata for a title, the date, and author declaration, and their access key (the study ID). %
The author declaration was a text field such that participants could list any entity they wanted.
Each of the postcards was composed with one of the four \ims{} (i.e., different levels of influence).
The order of the \ims{} was counterbalanced.
After all \ims{}, we presented a final questionnaire on the interaction experience, perceived control, and how they liked the resulting postcards. For questions on a specific \im{}, we included a screenshot of the corresponding UI.

\paragraph{Part 3: Text Recognition \& Rating}
The Prolific invitations for Part 3 were sent out two days after completing Part 2.
In the final questionnaire, participants were asked how well they remembered the postcard texts and what interaction method they had been created with. The delay was introduced to identify possible effects of text personalization on memory consolidation \cite{schone_experiences_2019}, as we also measured their text recognition performance.
Finally, we queried participants' opinions on automatic text generation and whether it should be mandatory to mark texts that were created with the help of artificial intelligence. We add the results for this part as supplementary material. 

\subsubsection{Apparatus \& Measures}

Study 1 was designed as a browser-based online experiment including three questionnaires, the custom interaction methods described in \autoref{sec:system_design}, and a custom blog-style website. This website included a landing page with example postcards (cf. \autoref{fig:screenshot_blog}) and an upload form to enter postcard texts and add metadata (cf. \autoref{fig:screenshot_upload}).
The components were coordinated with a study framework that handled counterbalancing, logging, and participant flow between the study components.
In the fine-tuning questionnaire, we recorded the texts and prompts the participants wrote.
There is yet no established methodology for measuring the sense of ownership, authorship, leadership, or the level of personalization. Consequently, in Part 2 and the final questionnaire (Part 3), we resorted to using custom-made visual analog scales as detailed below.

\paragraph{Authorship}
We measured the declaration of authorship. We operationalize this as the entity or entities participants listed as the credited author(s) in the blog upload form. In Study 1, this is done in an open format and motivated our choice for selection options in Study 2. This informs RQ1 and RQ4, addressing the perception and attribution of authorship in AI-generated texts. The measurement is used to test H1.2, H5, and H6.

\paragraph{Ownership}
The \textit{sense of ownership} indicates to what extent the participants felt as the authors of the text and, conversely, to what extent they felt that the AI was the author. ``To whom should this text belong'' was answered on a visual analog scale (range of -50 me and +50 AI). This measure informs RQ1 and RQ4 as well and is used to test our hypotheses on ownership (H1.1, H3, H4 \& H6).

\paragraph{Control}
With the \textit{sense of control}, we assessed whether participants felt they had control over the resulting text. This was measured with the item ``Who was in control over the content of the postcard?'', which was answered on a visual analog scale (range of -50 me and +50 AI). This relates to our RQ on control and is used to test our hypothesis on the effect of \ims{} on the sense of control testing H2.1.

\paragraph{Leadership}
The \textit{sense of leadership} denotes who participants felt took the lead during the writing process. It was tested with the item ``Who took the lead in writing?'' answered on a visual analog scale (range of -50 me to +50 AI) This informs RQ2 and allows us to test H2.2.

\paragraph{Personalization}
We evaluate two metrics of personalization. 
The \textit{objective quality of the personalization} provides a comparison between the texts used for fine-tuning and the generated texts with or without personalization.
The \textit{subjective quality of the personalization} addresses the participants' impression of how well the texts produced with the \ims{} matched their style. The latter informs RQ3 and tests H3.

\paragraph{User experience}
Lastly, we assess the \textit{user experience with the \ims{}} with the AttrakDiff questionnaire \cite{hassenzahl_inference_2010} and open-ended questions. The long version of the AttrakDiff assesses four aspects of user experience with antonymous adjective pairs: (1) the \textit{pragmatic quality} or usability of a product, its \textit{hedonic quality} or pleasantness in terms of (2) \textit{stimulation}, i.e., being inspiring and facilitating personal development, (3) \textit{identity}, i.e., communicating a desired identity, (4) and its overall \textit{appeal} \cite{hassenzahl_attrakdiff_2003,hassenzahl_inference_2010}. This provides insights into the \ims{} and future implications for designing user interfaces for LLMs and can highlight how user experience and control relate to approach RQ2.

\autoref{tab:study1_measures} provides the full list of measures for these constructs and for auxiliary measures (e.g., perceived creepiness for a broader perspective on user acceptance).
Demographic information was exported from Prolific.

\subsubsection{Participants}

We recruited 43 native English speakers via Prolific, of which 30 correctly completed all required steps of the study.
Seven participants withdrew from the fine-tuning questionnaire. Of the remaining 36 participants, one never started Part 2, and one had technical issues. A third participant was excluded because they took more than two days between the first and last step of Part 2, and this would have affected their answers in Part 3. Thus, we invited 33 participants to Part 3, and 32 completed it. Two of these uploaded duplicate texts, which meant their data had to be discarded. 
Of the final set of 30 participants, 13 identified as female and 17 as male, and none as non-binary or other. They were 37.2 years old on average ($ SD = 11.8 $, $ min = 20 $, $ max = 70 $). They listed the UK (17), the US (9), South Africa (2), Canada (1), and Zimbabwe (1) as their country of residence.
Five participants had no prior experience with text-generation systems. Nineteen (63\%) had used word or sentence suggestions, and 20 (67\%) had used auto-correction features on their devices. Half of the participants had previously written texts with auto-completion, and 8 (27\%) had used a smart-reply option. Additionally, two participants stated that other people sometimes write texts on their behalf, for example, an assistant at work. %
The participants received a compensation of £18 for Part 1, £13.50 for Part 2, £3 for Part 3, and a bonus payment of £6 for finishing all steps, i.e., a total of £40.50 for approximately four hours of their time.

\subsection{Results}

We first investigate whether the personalization of the texts was successful. We then turn to the analysis of the sense of ownership and the declaration of authorship (H1.1 and H1.2), including the effect of \textsc{Personalization} (H3). Finally, we examine whether the sense of control and sense of leadership were manipulated by the interaction method (H2.1 and H2.2). We expand this by analyzing the relationship between the sense of control and leadership with the sense of ownership.
An in-depth analysis of the remaining measures (including the open-ended questions) was beyond the scope of this paper. We provide a short overview of additional results via an OSF repository\footnote{\url{https://osf.io/n4svx/?view_only=916873e81d244be6be8e0790531b1197}}. In particular, the repository provides includes the text remembrance analysis, qualitative statements on the \ims{}, and an overview of additional measures linked to ownership, leadership, and control.

\paragraph{Analysis}
Below, we analyze differences between the two \textsc{Personalization} groups and four \ims{} regarding authorship, using mixed analysis of variance (ANOVA) with Bonferroni-corrected $t$-tests in post-hoc comparisons. For repeated-measures nominal data, we use Cochran's $Q$-test. For estimating regression coefficients in hierarchical repeated measures, we use linear mixed models (LMM) with Kenward-Roger estimation of the degrees of freedom. We use effect coding for two-level variables. To compare interaction methods in Study 1, we contrast each condition to the \writing{} condition. Non-parametric ordinal data is tested with a rank-aligned repeated-measures ANOVA \cite{kay_mjskayartool_2021}; here, follow-up analysis is based on Bonferroni-corrected ART-C contrasts \cite{elkin_aligned_2021}.
NULL-hypotheses relevant to our main research questions are evaluated using Bayes factors computed using the BayesFactor package \cite{BayesFactor}. %
The repeated-measures ANOVA model is robust to violations of normality with regard to their residuals \cite{glass1972consequences,harwell1992summarizing}. Thus, for the sake of brevity, the main text only reports the parametric model (and not also the non-parametric one) in cases where a d'Agostino normality test \cite{d1971omnibus} (Shapiro-Wilk test is too sensitive for $n >50 $) indicates normality violations and the results of non-parametric and parametric models align. We apply a Greenhouse-Geisser correction when the assumption of sphericity is violated.
For statistical inference, $\alpha$ was set at 5\%. We use the significance markers $ ^\ast $ for $ p < 0.05 $, $ ^{\ast\ast} $ for $ p < 0.01 $, and $ ^{\ast\ast\ast} $ for $ p < 0.001 $. 
Finally, we compute cosine similarity to compare texts using the tidytext package in R \cite{Silge2016}.

\paragraph{Manipulation Check: Personalization of the Generated Texts}

To check whether the personalization was successful, we computed the cosine similarity of each participant's concatenated responses from the fine-tuning survey and the concatenated AI-generated texts presented to the respective participant in Part 2. We expected that the similarity of texts for the true personalization group should be larger than for the placebo-personalization group that received non-personalized texts. Indeed, on average, the cosine similarity was larger for the true personalization group ($M = .77$, $SD = .04$) compared to the placebo-personalization condition ($M = .72 $, $SD =.06 $, $t(25.72) = -2.89$, $p = .008$, $d = -1.06$). 
\autoref{tab:personalization_getting} confirms that participants found the personalized texts tended to be a better match for the participants' personal writing styles than the placebo-personalized texts.

\begin{table*}
    \centering
    \caption{Participants' perception of the text with the interaction method \getting{} (1 Strongly disagree -- 7 Strongly agree)}
    \begin{tabularx}{\linewidth}{X *4{>{\centering\arraybackslash}p{.08\textwidth}}}
        \toprule
         & \multicolumn{2}{c}{\textbf{Personalized}} & \multicolumn{2}{c}{\textbf{Placebo-Personalized}}  \\ 
        \cmidrule(r){2-3} \cmidrule(r){4-5}
         & \textbf{\textit{MD}} & \textbf{\textit{SD}} & \textbf{\textit{MD}} & \textbf{\textit{SD}} \\ 
        \midrule
        The postcard mostly contains words and/or phrases that I usually use when writing in English. &
            5.07 & 1.67 &
            4.20  & 1.90 \\
        \midrule
        I would have written a similar postcard by myself. &
            4.40 & 1.99 &
            3.53  & 2.10 \\
        \bottomrule
    \end{tabularx}
    \label{tab:personalization_getting}
\end{table*}

\begin{table*}
    \centering
    \caption{Participants' responses to selected Likert matrix questions (1 Strongly disagree -- 7 Strongly agree)}
    \label{tab:S1_likert_measures}
    \begin{tabular}{p{.5\textwidth} | c c|c c|c c|c c}
        & \multicolumn{2}{c |}{\textbf{\writing{}}} & \multicolumn{2}{c |}{\textbf{\editing{}}} & \multicolumn{2}{c |}{\textbf{\choosing{}}} & \multicolumn{2}{c}{\textbf{\getting{}}} \\
        & \textbf{\textit{MD}} & \textbf{\textit{SD}} & \textbf{\textit{MD}} & \textbf{\textit{SD}} & \textbf{\textit{MD}} & \textbf{\textit{SD}} & \textbf{\textit{MD}} & \textbf{\textit{SD}} \\
        \midrule
        1: I felt like the AI was acting as a tool which I could control.
            & 4 & 2.15 & 5 & 1.52 & 5 & 1.75 & 4 & 2.27 \\
        \midrule
        2: I felt like the AI system was acting as a ghostwriter, writing the postcard on my behalf.
            & 2.5 & 2.18 & 5 & 1.63 & 6 &1.71 & 6.5 & 2.32 \\
        \midrule
        3: I felt like I was writing the text and the artificial intelligence was assisting me.
            & 1.5 & 2.44 & 3 & 1.99 & 2.5 & 2.16 & 2 & 2.03 \\
        \midrule
        4: I felt like the artificial intelligence was writing the text and I was assisting.
            & 1 & 1.73 & 5.5 & 1.91 & 5 & 1.82 & 4 & 2.13 \\
        \bottomrule
    \end{tabular}
\end{table*}

\begin{figure}
    \centering
    \includegraphics[width=\columnwidth]{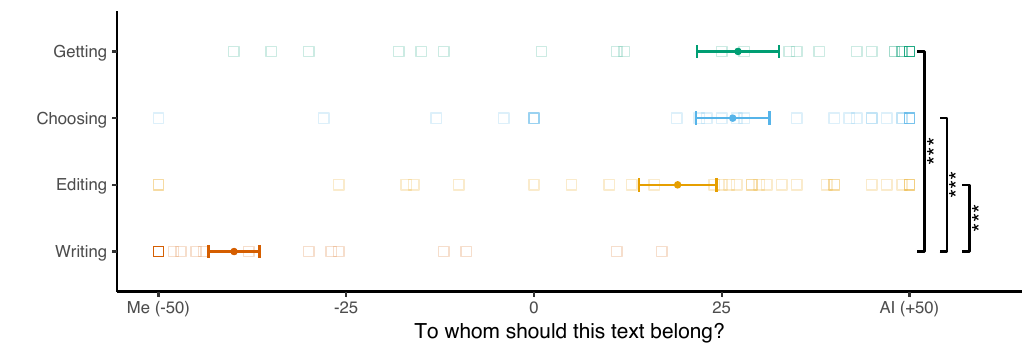}
    \caption{Mean ratings for each \im{} of the item ``To whom should this text belong'' answered on a visual analog scale (range of -50 me and +50 AI) with individual data points as boxes.  Error bars indicate $ \pm $ one standard error of the mean.}
    \label{fig:S1_belong}
\end{figure}

\paragraph{H1.1 \& H3: Sense of Ownership}
\begin{table}
    \centering
    \caption{Repeated-measures ANOVA results for \im{}, \textsc{Personalization}, and their interaction. When the sphericity assumption was violated, we applied a Greenhouse-Geisser correction to all within-subject factors. Corrected results are marked with a $ ^G $. Significance markers: $ ^{\ast\ast} $ for $ p < 0.01 $, $ ^{\ast\ast\ast} $ for $ p < 0.001 $ }
    \begin{tabularx}{\linewidth}{X l c c c c}
        \toprule
        \textbf{Measure} & \textbf{Effect} & $F$ & $ df $ & $ p $ & $\eta^2_p$ \\
        \midrule
        \multirow[t]{3}{=}{Sense of ownership: To whom should this text belong?} &
            \im & $55.75^G$ & $(2.02, 56.58)$ & $\bold{< .001}^{\ast\ast\ast}$ & $.67$ \\
        & \textsc{Personalization} & $0.52$ & $(1,28)$ & $.477$ & $.02$ \\
        & \textsc{IM.} $ \times $ \textsc{Pers.} & $0.48^G$ & $(2.02,56.58)$ & $.625$ & $.02$ \\
        
        \multirow[t]{3}{=}{Sense of control: Who was in control over the content of the postcard?} &
            \im & $28.18^G$ & $(2.25, 62.95)$ & $\bold{< .001}^{\ast\ast\ast}$ & $.50$ \\
        & \textsc{Personalization} & $1.05$ & $(1,28)$ & $.315$ & $.04$ \\
        & \textsc{IM.} $ \times $ \textsc{Pers.} & $0.60^G$ & $(2.25,62.95)$ & $.570$ & $.02$ \\

        \multirow[t]{3}{=}{Sense of leadership: Who took the lead in writing?} &
            \im & $28.18^G$ & $(2.25,62.95)$ & $\bold{< .001}^{\ast\ast\ast}$ & $.50$ \\
        & \textsc{Personalization} & $1.05$ & $(1,28)$ & $.315$ & $.04$ \\
        & \textsc{IM.} $ \times $ \textsc{Pers.} & $0.60^G$ & $(2.25,62.95)$ & $.570$ & $.02$ \\

        \multirow[t]{3}{=}{AttrakDiff: pragmatic quality} &
            \im & $4.20$ & $(3,84)$ & $\bold{.008}^{\ast\ast}$ & $.13$ \\
        & \textsc{Personalization} & $1.85$ & $(1,28)$ & $.185$ & $.11$ \\
        & \textsc{IM.} $ \times $ \textsc{Pers.} & $3.46$ & $(3,84)$ & $.020$ & $.11$ \\

        \multirow[t]{3}{=}{AttrakDiff: hedonic quality - identity} &
            \im & $8.50^G$ & $(2.23,62.32)$ & $\bold{< .001}^{\ast\ast\ast}$ & $.23$ \\
        & \textsc{Personalization} & $0.76$ & $(1,28)$ & $.391$ & $.03$ \\
        & \textsc{IM.} $ \times $ \textsc{Pers.} & $2.24^G$ & $(2.23,62.32)$ & $.109$ & $.07$ \\

        \multirow[t]{3}{=}{AttrakDiff: hedonic quality - stimulation} &
            \im & $4.61$ & $(3,84)$ & $\bold{.005}^{\ast\ast}$ & $.14$ \\
        & \textsc{Personalization} & $0.34$ & $(1,28)$ & $.562$ & $.01$ \\
        & \textsc{IM.} $ \times $ \textsc{Pers.} & $2.29$ & $(3,84)$ & $.084$ & $.08$ \\

        \multirow[t]{3}{=}{AttrakDiff: appeal} &
            \im & $8.07$ & $(3,84)$ & $\bold{< .001}^{\ast\ast\ast}$ & $.22$ \\
        & \textsc{Personalization} & $0.97$ & $(1,28)$ & $.332$ & $.03$ \\
        & \textsc{IM.} $ \times $ \textsc{Pers.} & $2.27$ & $(3,84)$ & $.086$ & $.08$ \\
        
       \bottomrule
    \end{tabularx}
    \label{tab:s1_anova_results}
\end{table}

Next, we tested whether the applied \ims{} changed the sense of ownership for the text generated in the human-AI interaction. In a repeated-measures ANOVA with the factors \textsc{Personalization} (between, 2 levels; true personalization vs. placebo-personalization) and \im{} (within, 4 levels; \writing{}, \getting{}, \choosing{}, \editing{}), we analyzed the item ``To whom should this text belong'' that participants answered on a visual analog scale (range of -50 me and +50 AI)\footnote{The assumption of normality of residuals was violated ($p$ <.001). Therefore, we computed a non-parametric ANOVA \cite{kay_mjskayartool_2021}. As with the parametric model, there was an effect of 
\im{}, $F(3,84) = 45.24$, $p < .05$. Post-hoc tests resembled the parametric model; all other effects were $ p > .05$}. Due to the violation of the sphericity assumption, $W > .45$, $p < .001$, we applied a Greenhouse-Geisser correction to all within-subject factors.

As shown in \autoref{tab:s1_anova_results}, we found a significant effect of \im{}. All AI-related \ims{} differed from the \writing{} condition (all $p$ <.001), but none of the AI-related \ims{} differed from each other (all $p >.05$), see also \autoref{fig:S1_belong}. We did not find an effect of \textsc{Personalization}, or an interaction effect of \im{} $\times$ \textsc{Personalization}. We followed up on the non-significant main effect of \textsc{Personalization} with an independent Bayesian $t$-test and found that the model assuming no difference is 2.38 times more like than the model assuming differences in means given the priors\footnote{Priors were set to default here with $\sqrt{2}/2$ on the standardized Cauchy prior as implemented in \cite{BayesFactor}. Prior sensitivity checks showed that with wider priors the model assuming no difference between means became more likely.}. 

Therefore, we can follow that text generated by a (placebo-)personalized AI changed the sense of ownership (H1.1). In particular, all AI-related methods elicited a sense of ownership different from \writing{}. However, the personalization did not affect whether participants attributed authorship (H3).
Note that in \autoref{fig:S1_belong}, one can see that some participants attributed a sense of ownership to the AI in the \writing{} condition. This could be due to participants taking over formulations of the AI-generated texts from the other conditions or due to careless responses.

\paragraph{H1.2: Declaration of Authorship}

Did the sense of ownership correlate with declaring AI as an author on the website?
For this, we considered the free-text entries in the \textit{author} field of our blog upload form (see \autoref{sec:procedure_s1}). We classified the entries as referring to the participants themselves (i.e., a name or acronym), the AI, an impersonal attribution (``Human'', ``Random''), or a combination of several categories (e.g., ``AI and Kyle'').

\begin{figure}
    \centering
    \includegraphics[width=\columnwidth]{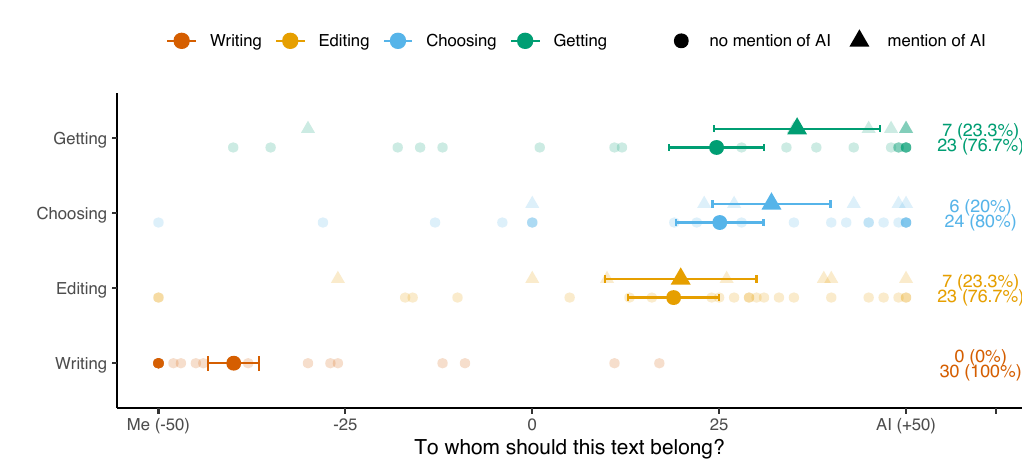}%
    \caption{Sense of ownership (x-axis) and declaration of authorship (marker shape: circle vs triangle) as a function of \im{} for AI-generated texts.  Error bars indicate $ \pm $ one standard error of the mean. The numbers on the right indicate the counts of authorship declarations binarized for any mention of AI as an author as a function of \im{} in our sample ($n = 30$). Note that all participants underwent all \ims{}. One participant declared the writer of the postcard to be ``random'' for all \ims{} that involved AI-generated texts. These were allocated to the no-mention-of-AI category.}%
    \label{fig:S1_declare_belong}
\end{figure}

Curiously, although participants had little sense of ownership, they still added their name to the postcard when publishing it on the website, see \autoref{fig:S1_declare_belong}. Note, however, that our results can only approach this from a correlational perspective.
We find that people declared themselves as authors for the \writing{} condition but that they also declared themselves as authors for the (placebo-)personalized AI-generated texts. We tested the differences in proportions against chance with Cochran's $Q$-test. \im{} had a significant effect on the frequency of mentioning ``AI'', $Q(3) = 18.54 $, $p < .001 $. The counts on the right side of \autoref{fig:S1_declare_belong} summarize how often AI was (not) listed when declaring authorship for the postcard. Only between six and seven participants per \im{} (except \writing{}) mentioned AI. For the \editing{} condition, all mentions of the AI were in collaboration with the participant or an impersonal attribution. For example, they put their name and ``AI''. For the \choosing{} condition, three of six mentions only listed ``AI''; the other three were declared as a collaboration of the AI and the participant or a ``human''. For the \getting{} condition, this was the case for only one of the seven occurrences; all other cases mentioned ``AI'' as the sole author.
Planned comparisons between the \writing{} and the other conditions were all significant (all $p <.03 $), however, no pairwise comparisons for \editing{}, \getting{}, \choosing{}, were significant (all $p >.05$). Thus, we follow that while AI was declared as an author for postcards produced with AI interactions, this was not the case for all participants (partial support for  H1.2). %

\autoref{fig:S1_declare_belong} further illustrates the relation between the sense of ownership and the declared authorship, contrasting the sense of ownership against the count data for all the conditions with AI-generated personalized texts. Although participants judged the text to belong to the AI in most of the cases, they still put their name underneath the postcard in most of the cases. We refer to this discrepancy as the \aighostwriter{}.

The questionnaire results and participant statements provide insights into reasons for the \aighostwriter. For example, when participants were not writing their own texts, they tended to feel that the AI was a tool and that it acted as a ghostwriter (see \autoref{tab:S1_likert_measures}). The level of collaboration between a participant and the AI (Question 3) was lowest for \writing{} and \getting{}, i.e., the conditions with no AI support and full AI support.
Seventeen participants (56.7\%) stated that, in their opinion, it should not be mandatory to mark texts that were created with the help of AI as such. Reasons they mentioned include that the AI \textit{``is really just a tool to say what the writer wants to say,'' and that ``as long as the person using the AI reviews the text before it is sent then it should be ok.''}
Several participants also felt that disclosing AI support would not be valued, e.g.: \textit{``People would feel offended if they thought I couldn't be bothered to text time myself.''}
Others explicitly mentioned human ghostwriters as a comparison: \textit{``prior to machine generated text there was a whole industry of humans filling content space in a similar manner.''}
The (placebo-personalization also made a difference: One participant from the placebo-personalized group noted that marking is not necessary, \textit{``especially if it reflects who you are and you chose the specific message that suits you the best.''}

Conversely, 13 participants (43.3\%) felt that disclosing AI contributions should be mandatory. Arguments included ethics and transparency, e.g.: \textit{``It's important to know whether you are interacting with a human or an algorithm''} and \textit{``to understand if something has been said that doesn't really make sense.''} Others mentioned self-protection and that otherwise \textit{``it feels a bit like cheating.''}

\paragraph{H2.1: Sense of Control}

\begin{figure}
    \centering
    \includegraphics[width=\columnwidth]{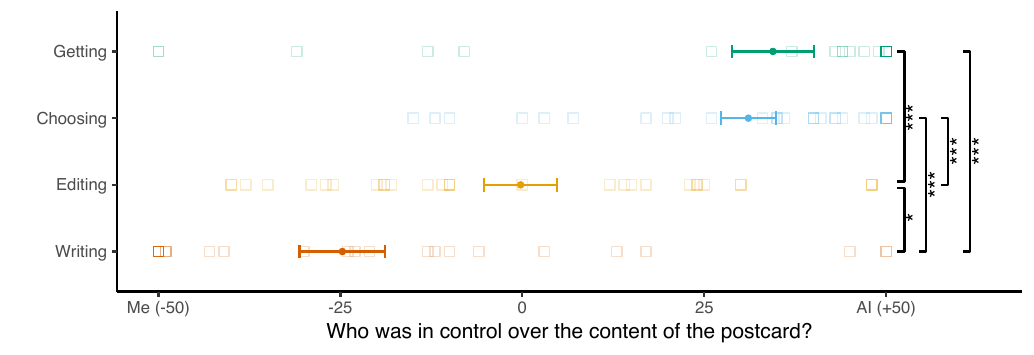}%
    \caption{Sense of control as a function of \im{} for AI-generated texts. Error bars indicate $ \pm $ one standard error of the mean.}%
    \label{fig:S1_control}
\end{figure}

We analyzed participants' responses on the item ``Who was in control over the content of the postcard?'', which was answered on a visual analog scale (range of -50 me and +50 AI) with a repeated-measures ANOVA model. %
Again the sphericity assumption was violated, $W > .60 $, $p < .02 $, and we applied a Greenhousse-Geisser correction to all within-subject factors and their interaction terms. The normality of residuals was not violated ($p = .14$). We found that \im{} affected the sense of control. Bonferroni-corrected $t$-tests revealed significant differences between all comparisons (all $p <.05 $), with the exception of the comparison between \getting{} and \choosing{} ($p > .99 $, see also 
\autoref{fig:S1_control}). 
Therefore, we can consider that our manipulation of \im{} successfully varied the sense of control in the human-AI interaction. 

As shown in \autoref{tab:s1_anova_results}, there was neither an effect of \textsc{Personalization}, nor an interaction effect. Comparing personalization with a Bayesian $t$-test\footnote{Priors were set to default here with $\sqrt{2}/2$ on the standardized Cauchy prior, as implemented in \cite{BayesFactor}. Again, prior sensitivity analysis indicated that with wider priors, the model assuming no difference between means became more likely.}, we could quantify that the model assuming no difference was 1.98 times more likely than the model assuming differences between groups.

\begin{figure}
    \centering
    \includegraphics[width=.90\columnwidth]{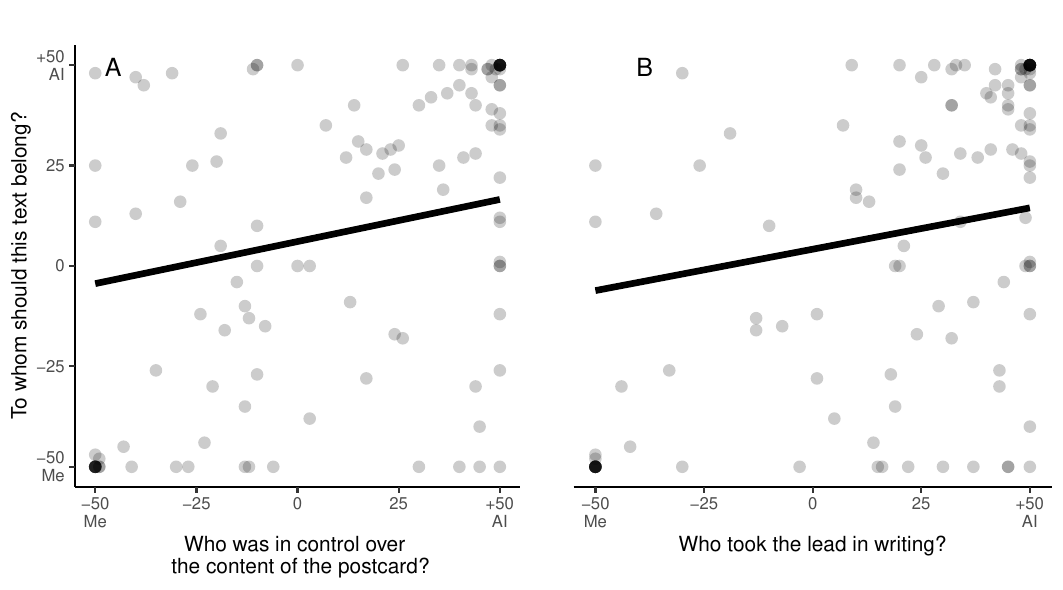}%
    \caption{A: Sense of control as a function of the sense of ownership for AI-generated texts. B: Sense of leadership as a function of the sense of ownership for AI-generated texts. Dots show the raw data, and the line represents the marginal effect of the sense of control on the sense of authorship without the effect of \im.}%
    \label{fig:S1_control_authorship}
\end{figure}

We expanded the ANOVA model by fitting a linear mixed model to predict the sense of ownership from the sense of control and the \im{}. The model included participants as a random effect to account for the structure of the repeated measures in the data. The model's total explanatory power was substantial (conditional $R^2 = 65\%$; marginal $R^2 = 55\% $). As in the prior analysis of authorship, we found a significant effect of \im{} on the sense of ownership, $F(3,86.02)=28.00 $, $ p <.001 $. The sense of control also predicted the sense of ownership, $ F(1,101.55)=6.47 $, $ p = .02$. There was a statistically significant and positive effect ($\beta = 0.21 $, 95\% CI [0.05, 0.37], $t(109) = 2.57$, $p = 0.012$;  $\beta_z = 0.20$, 95\% CI [0.05, 0.36]). For a predictive plot of the effect of sense of control on the sense of ownership with raw data, see \autoref{fig:S1_control_authorship}A. Neither the main effect of \textsc{Personalization} nor the interaction of \textsc{Personalization} $\times$ \im{} was significant (both $p > .05$).

\paragraph{H2.2: Sense of Leadership}

\begin{figure}
    \centering
    \includegraphics[width=\columnwidth]{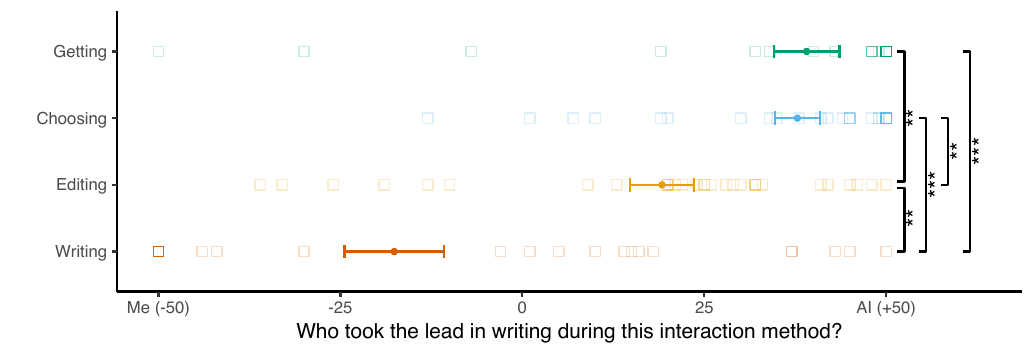}%
    \caption{Sense of leadership as a function of \im{} for AI-generated texts. Error bars indicate $ \pm $ one standard error of the mean.}%
    \label{fig:S1_lead}
\end{figure}

Participants' responses to ``Who took the lead in writing?'', again answered on a visual analog scale (range of -50 me to +50 AI), were analyzed in a mixed ANOVA model\footnote{Again the assumption of normality of residuals was violated ($ p = 0.03$). We, therefore, computed a non-parametric ANOVA \cite{kay_mjskayartool_2021}. Resembling the parametric model, there was an effect of \im{}, $F(3,84) = 30.26$, $p < .05$. Non-parametric post-hoc tests resembled the parametric analysis.}. As shown in \autoref{tab:s1_anova_results}, there was a significant effect of \im{}, but no effect of \textsc{Personalization}, nor an interaction effect. Note that a Bayesian $t$-test, as modeled for the sense of control, indicated that the model assuming no difference for \textsc{Personalization} was 2.52 times more likely than the model assuming differences. Post-hoc tests showed significant differences for all \ims{} with the exception of the difference between \getting{} and \choosing{}, $p > .99$, see \autoref{fig:S1_lead}.

We again expanded the ANOVA by fitting a linear mixed model to predict the sense of ownership from the sense of leadership. We found a similar pattern of results. There was a significant effect of \im{} on the sense of ownership, $F(3,84.91)=23.71$, $ p <.001 $ and no main or interaction effect with \textsc{Personalization} (both $p$ >.05) but a significant main effect of sense of leadership, $F(3,95.65)=5.93$, $ p =.016 $. The relation between the sense of ownership and the sense of leadership was positive 
($\beta = 0.21 $, 95\% CI [0.04, 0.37], $t(109) = 2.46$, $p = 0.015$;  $\beta_z = 0.19$, 95\% CI [0.04, 0.34]), see also \autoref{fig:S1_control_authorship}B.
The model's total explanatory power was substantial (conditional $R^2 = 64\%$; marginal $R^2 = 55\% $).

\ims{} affected the sense of control over the content and sense of leadership in the interaction. This predicted the sense of ownership and, thus, the feeling to whom a text should belong (H2).

\subsection{User Experience}
We analyzed all subscales of the AttrakDiff with a repeated-measures ANOVA model with Greenhouse-Geisser correction. All ANOVA results are shown in \autoref{tab:s1_anova_results}.

For \textit{hedonic quality -- identification} there was no effect of \textsc{Personalization}, but a significant effect of \im{}. Bonferroni-corrected post-hoc comparisons on \im{} indicated that the \getting{} ($M = 3.74 $, $SD =1.28 $) interaction was perceived worse than \writing{} ($M = 4.78 $, $SD = 0.88 $), \choosing{} ($M = 4.47 $, $SD = 1.21$), and \editing{} ($M = 4.81 $, $SD =1.06 $; all other comparisons $p>.05$). The \textsc{Personalization} $\times$ \im{} term did not reach significance. Therefore, the \getting{} interaction can be regarded as the least self-relatable interaction method interaction.

For the ANOVA on \textit{hedonic quality -- stimulation}, we find a similar pattern of results, with significant differences for \ims{} but not for \textsc{Personalization} and the interaction \textsc{Personalization} $\times$ \im{}. Comparisons that were Bonferroni-corrected on the factor \im{} only indicated differences between the \getting{} ($M = 3.91 $, $SD = 1.33 $),  and the \editing{} ($M  = 4.81 $, $SD = 1.06 $) condition (all other comparisons $ p>.05 $). Unsurprisingly, the \getting{} interaction was, thus, deemed to be less stimulating than the \editing{} method.

For the analysis of attractiveness, the same pattern applies. Again, \getting{} ($M = 3.99 $, $SD = 1.66 $),  was significantly less attractive than \writing{}, ($M = 5.24 $, $SD = 0.89 $), \choosing{} ($M = 4.94 $, $SD = 1.39 $), and \editing{} ($M = 5.13 $, $SD =1.38 $; all other comparisons $p >.05 $).

The ANOVA on the pragmatic qualities revealed a different pattern of results. There was again no effect of \textsc{Personalization} but an effect of \im{}. Note, however, that this was qualified by a significant \textsc{Personalization} $\times$ \im{} interaction term. A Bonferroni-corrected comparison of means across \textsc{Personalization} $\times$ \im{} revealed that pragmatic qualities for the participants that did not receive personalized texts in the \getting{} condition ($M = 4.32 $, $SD = 1.04 $) significantly differed from the \choosing{} condition ($M = 5.14 $, $SD = 0.84 $) but also from the \writing{} condition ($M = 5.70 $, $SD = 0.74 $; all other comparisons $p >.05 $) in the group that received personalized texts. This indicates that \getting{} is not subjectively perceived as very pragmatic, even though it is clearly the \im{} with the most time-efficient interaction pattern.

\begin{figure}
    \centering
    \includegraphics[width=.5\textwidth]{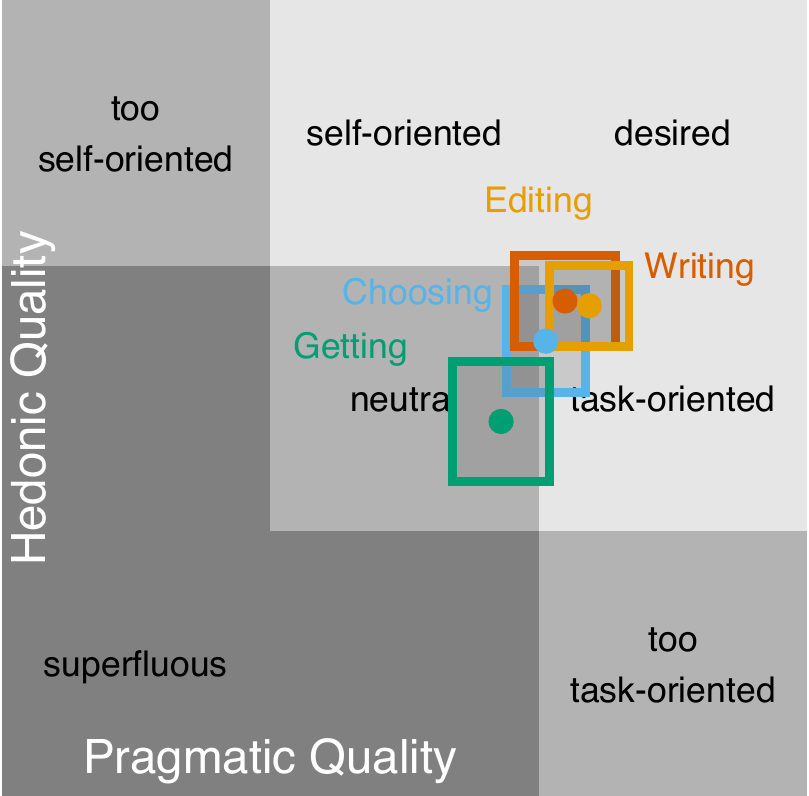}
    \caption{AttrakDiff results for the \ims{} in Study 1. The graphic shows the mean (dot) and standard error (rectangles) of the aggregated AttrakDiff scores in the two dimensions \textit{Pragmatic Quality} and \textit{Hedonic Quality}.}
    \label{fig:s1_attrakdiff}
\end{figure}

Overall, \autoref{fig:s1_attrakdiff} shows that \getting{} was the most neutral \im{} in absolute terms, and that the user experience for \writing{} and \editing{} was comparable. \choosing{} used minimal interactions with the AI model and was, thus, probably rated as more pragmatic than \getting.

\section{User Study 2: Declared Authorship with Human and AI Ghostwriters}

With Study 1, we demonstrated that personalized AI-generated texts diminish the sense of ownership compared to self-written texts. This stands in contrast to the declaration of authorship. In the majority of cases, participants did not declare the AI system as an author of the text when publishing it online. We summarized this as the \aighostwriter{}. We found that \ims{} affect the sense of control in the human-AI interaction, which increases the sense of ownership. We found that \textsc{Personalization} did not increase the sense of ownership or sense of control, or descriptively, authorship declaration. Therefore, true personalization was not a necessary condition for the \aighostwriter{}.
This pattern of results allows us to run a Wizard-of-Oz study with respect to personalization. Therefore, in Study 2, we describe the postcards to be personalized but actually assign the non-personalized texts randomly to the participants. 

In Study 2, we address four important limitations of the previous study: 
First, we found that participants attributed a sense of authorship to the AI for the self-written texts. This could be due to our experimental design where people underwent all \ims{} and only then were asked to declare their sense of ownership for the four postcards. Participants could have taken up information in their self-written texts that appeared in the previous AI-generated texts.
We address this limitation by only using one \im{}. To thoroughly test the \aighostwriter{}, we chose \getting{}, because it had the lowest sense of ownership and the most mentions of AI as an author.
Second, to attribute any authorship to the AI-generated texts, participants had to come up with a label for the AI. This made one participant label the postcard with ``random''.
In Study 2, we provide participants with six pre-defined authorship labels that they can choose from, motivated by Study 1.
Third, one can entertain that not declaring collaboration is not specific to human-AI interaction but also happens in human-human interaction, i.e., in the case of human ghostwriters. Thus, we add a fictional human ghostwriter.
Fourth, the small sample size did not allow for the accurate estimation of relative frequencies in declaring the AI as an author. A larger study will allow for a more accurate estimation of this effect. As listed in \autoref{sec:research_questions}, we hypothesize that we can replicate the \aighostwriter{} and that it exceeds the ghostwriter effect for human-human interaction. 
Study 2 was conducted in December 2022.

\subsection{Method}

\subsubsection{Procedure}

The study consisted of two parts: (1) a placebo-personalization survey and (2) interaction with generated postcards. 
Before starting with Part 1, we informed participants about the study procedure and their rights and obtained consent.

\paragraph{Part 1: Personalization Questions}
First, participants were asked to upload five texts of at least 500 words that they had written in the past. We told them that the texts should preferably be informal texts such as personal emails and that they had the option to remove or change any detail they did not want to share. This part was intended to make participants believe that texts would be generated for them.
Since the previous study had shown no effect of personalization, the texts were not actually used.  However, participants entered a name or pseudonym, and we performed a placebo-personalization by applying the provided name as the greeting in the postcard.
Participants also answered questions on their attitude towards AI (cf. \autoref{tab:study2_measures}).

\paragraph{Part 2: Authorship Attribution}
One day later, we randomly presented participants with one of the 15 non-personalized AI-generated postcard texts used in Study 1. They were told that this text was written either by a human or by an AI ghostwriter and that the text was personalized using their input from Part 1.
They were then shown a blog-style upload form for the generated text and were asked to choose an image and provide metadata: a title, the date, and an author attribution. This time, the author(s) were selected from a randomized list including the options \textit{AI}, \textit{GPT-3}, \textit{Human}, \textit{Sasha} (the name of the fictional ghostwriter), \textit{Other + text field}, and the \textit{participants' name}, \textit{initials}, or \textit{pseudonym} provided in Part 1.
Directly after the upload, the participants answered questions on their perceived agency and how they liked the resulting postcard.
This process was repeated for the second ghostwriter type, i.e., AI or human.
The order of the conditions was counterbalanced.
After the interaction with the two ghostwriter types, we administered questions on the participants' reasoning on authorship attribution and re-assessed their attitude towards AI after participation in the study.

\subsubsection{Apparatus \& Measures}

The study was implemented with two Qualtrics surveys.
As in Study 1, we measured the participants' sense of ownership, control, and leadership. Moreover, we counted how often AI was mentioned as an author.
An overview of all recorded measures is shown in \autoref{tab:study2_measures}.

\subsubsection{Participants}

We recruited 107 native English speakers via Prolific. One participant was excluded after Part 1 for not following the instructions. All others were invited to participate in Part 2, and 100 did so. Of these, four failed an attention check. Analyses only include the remaining 96 participants that correctly completed both parts. These reported their gender as male (53) or female (43); the options non-binary, other, or undisclosed were not selected. They were between 20 and 75 years old ($ M = 36.3 $, $ SD = 13.3 $). Their current countries of residence were the UK (41), South Africa (25), Canada (15), Ireland (5), and 1-3 participants each listed Mexico, Australia, Spain, Israel, Portugal, and the US.
As in Study 1, we asked for their experience with text generation. Fifty-five said they had previously used smart-reply features, 81 word or sentence suggestions, 63 auto-completion, and 86 auto-correction. Thirteen participants sometimes have people writing texts on their behalf.
The full study took around 40 minutes to complete, and participation was compensated with £5. %

\subsection{Results}

This section evaluates hypotheses 4-6 and follows up with a correlational analysis of the sense of control and leadership. The OSF repository\footnote{\url{https://osf.io/n4svx/?view_only=916873e81d244be6be8e0790531b1197}} provides additional results on measures linked to the sense of control, user experience, and the participants' attitudes toward AI.

\paragraph{Analysis}
Below, we analyze differences between the two \textsc{Ghostwriter} groups regarding ownership using paired $t$-tests. When a d'Agostino test \cite{d1971omnibus} indicates a violation of the normality assumption, we also compute non-parametric Wilcoxon tests.
As for Study 1, NULL-hypotheses relevant to our main research questions are evaluated using Bayes factors computed with the BayesFactor package \cite{BayesFactor}.
We compared the frequency of adding a human author label (\textit{Sasha}/\textit{Human}) to texts presented as human-written versus adding an AI author label (\textit{GPT-3}/\textit{AI}) to AI-written texts with a McNemar $ \chi^2 $ test.
We applied Spearman rank correlation to evaluate the correlation between the sense of ownership and authorship declarations ($ 0 =$ no mention of \textsc{Ghostwriter}, $ 1 =$ mention of \textsc{Ghostwriter}).
Again, the significance level $ \alpha $ was set at 5\%.

\begin{table}
    \centering
    \caption{Welch-corrected $ t $-test results comparing human and AI \textsc{Ghostwriters}, including Cohen's $ d $ for effect sizes. When the normality assumption was violated, we also applied a non-parametric Wilcoxon test. Significance marker: $ ^{\ast} $ for $ p < 0.05 $}
    \begin{tabularx}{\linewidth}{X c c c c c c}
        \toprule
        \textbf{Measure} & \multicolumn{3}{c}{\textbf{Welch-corrected $ t $-test}} & \multicolumn{3}{c}{\textbf{Wilcoxon test}} \\
        & $ t(95) $ & $ p $ & $ d $ & $ W $ & $ p $ & $ r $ \\
        \midrule
        Sense of ownership: To whom should this text belong? &
            $-2.05$ & $\bold{.043}^\ast$ & $ -0.21$ &
            $1095 $ & $\bold{.018}^\ast$ & $.10 $ \\
        Sense of control: Who was in control over the content of the postcard? &
            $ 0.28$ & $.784$, & $0.03$ & & & \\
        Sense of leadership: Who took the lead in writing? &
            $ -1.10$ & $.274$ & $ -0.11$ &
            $ 876 $ & $\bold{.021}^\ast $ & $r = .12 $ \\
        \bottomrule
    \end{tabularx}
    \label{tab:s2_test_results}
\end{table}

\begin{figure}
    \centering
    \includegraphics[width=\columnwidth]{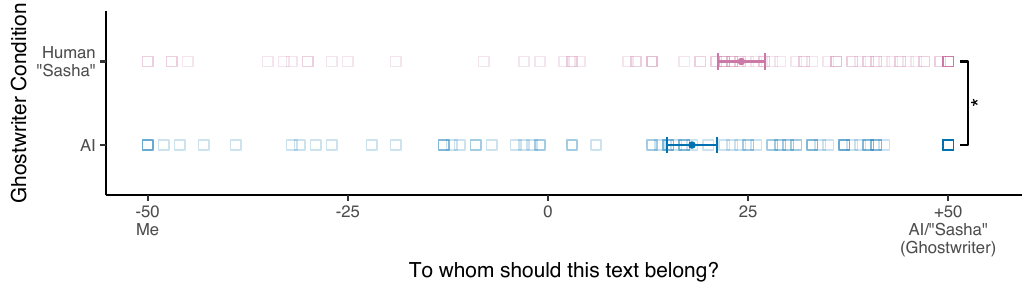}
    \caption{Sense of ownership as a function of \textsc{Ghostwriter}. Error bars indicate $ \pm $ one standard error of the mean.}
    \label{fig:s2_belong}
\end{figure}

\paragraph{H4: Differences in Sense of Ownership}
We compared the sense of ownership for the \human{} and \ai{} condition and found a significant difference, cf. \autoref{tab:s2_test_results}. A non-parametric Wilcoxon test was also significant (normality was violated, $p<.01$). People attributed more ownership to themselves when the ghostwriter was presented as an AI, compared to when it was presented as a human, see \autoref{fig:s2_belong}.

\paragraph{H5: Differences in Authorship Declaration}
We found differences in the declaration of authorship (H3). Generally, while people more often mentioned AI as an author (about 71.3\% in Study 2 versus a max. of 23.3\% in Study 1; we will discuss this in \autoref{sec:limitations}), we could see that they mentioned AI significantly less often, $\chi^2(1) = 31.92$, $p<.001$, than mentioning ``Sasha'' or any entity other than themselves, see \autoref{tab:S2_count_data}.

This trend also shows in the question of whether it should be mandatory to mark texts that were created with the help of a human/AI as such. For the human condition, 63 participants (65.6\%) answered \textit{yes}, compared to 59 (61.5\%) for the AI condition (all others selected \textit{no}).
For the declaration of human contributions, participants listed reasons such as emotional involvement, e.g., \textit{``we don't want to hurt their feelings, right?''} and \textit{``I would expect to be given credit if it was me,''} and that \textit{``it is their intellectual property.''}
Another argument was derived from the perceived authenticity of human-written texts, e.g., \textit{``if it's truly from a human telling you about the lived experience then it should have their name on it''} and \textit{``because it truly reflects the value of a human being.''}
One participant who felt that only human ghostwriters should be credited explained that \textit{``that person is not [a] tool, whereas the system AI is a tool that can [be] used.''}
Reasons for not naming a human ghostwriter were references to existing ghostwriting practices, e.g., \textit{``people that do this for a living''} and \textit{``it is the equivalent of a personal assistant, who wouldn't necessarily sign their own name.''} One participant again referred to the personalization: \textit{``i am a little bit in between but as i had given examples of my writing style then no.''} 

As in Study 1, participants argued that AI contributions should be credited for ethics and transparency, e.g.: \textit{``It seems ethical to declare this''}, and it \textit{``help[s] give an idea to the receiver to understand the background context.''}
In addition, not declaring contributions \textit{``could have legal ramifications.''}
Interestingly, several participants added that crediting AI confirms its potential, e.g., \textit{``this will also help put a positive image on AI.''} and \textit{``It gives people an understanding of how far AI has come.''}
Those who did not consider marking AI mandatory frequently described AI as a tool, e.g., \textit{``it expresses the feelings and viewpoints of the owner. it is just a tool used by the owner.''} and \textit{``AI is a tool we can use, just like we don't mark that we used Word or Google Docs to show we made something.''}
Similarly, one person said that \textit{``Artificial Intelligence is used on behalf of a person. It should be up to the person publishing whether or not to disclose AI usage.''}
The context also played a role, e.g., \textit{``the recipient might think that not much thought would have been sent into the sending of the postcard.''}

\paragraph{H6: \aighostwriter{}}
To evaluate H6, we computed the Spearman rank correlation of the sense of ownership and declaration of AI as an author. The link between the sense of ownership and declaring authorship of AI was significant but very small, $r_s = .24$, $p = .017$. Note that the same holds true for the link in human ghostwriters, $r_s = .26$, $p = .011$. Therefore, there is no direct 1:1 mapping between the sense of ownership, and there is no strong correlation, supporting H6.

\begin{figure}
    \centering
    \includegraphics[width=\columnwidth]{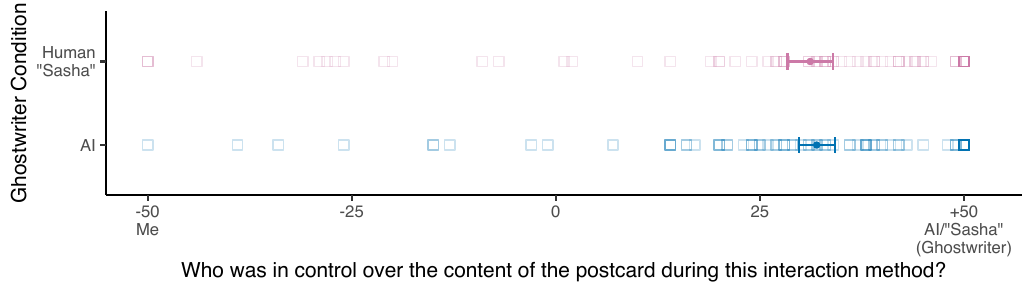}
    \caption{Sense of control as a function of \textsc{Ghostwriter}. Error bars indicate $ \pm $ one standard error of the mean.}
    \label{fig:s2_control}
\end{figure}

\begin{figure}
    \centering
    \includegraphics[width=\columnwidth]{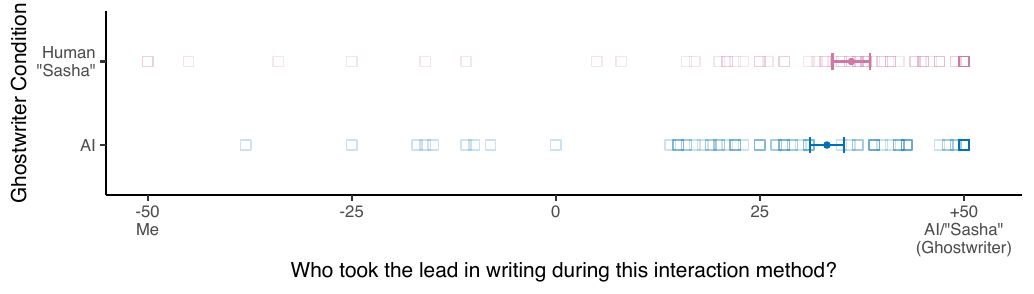}
    \caption{Sense of leadership as a function of \textsc{Ghostwriter}. Error bars indicate $ \pm $ one standard error of the mean.}
    \label{fig:s2_lead}
\end{figure}

\begin{table}%
\caption{Counts for each Ghostwriter on declared authorship binarized for any mention of AI/Human as authors. Note that all participants underwent all \textsc{Ghostwriter} conditions.}
\begin{tabular}{lcc}

                  &AI & Human\\ \hline
mention of Ghostwriter    & 71 (73.9\%)       & 79 (82.3\%)      \\
no mention of Ghostwriter & 25 (26.0\%)     & 17 (17.7\%)    \\
\end{tabular}
\label{tab:S2_count_data}
\end{table}

\paragraph{Sense of Control and Leadership}
We found no difference in sense of control between both conditions\footnote{A Bayesian $t$-test with default priors indicated that the model assuming no differences is about 8.5 times more likely than the model assuming a difference.}, cf. \autoref{tab:s2_test_results} and \autoref{fig:s2_control}.
For the sense of leadership, we found no significant difference in means with a Welch-corrected $t$-test, cf. \autoref{tab:s2_test_results}. However, there was a significant effect when comparing ranks with a Wilcoxon test, which was probably due to the large skew in the data (normality was violated, $p<.01 $). As shown in \autoref{fig:s2_lead}, the participants' sense of leadership was higher for the ghostwriter presented as an AI.

\section{Discussion}

Our studies show that there is a discrepancy between the sense of ownership and declaration of authorship when working with personalized AI text generation. This constitutes the \aighostwriter{}. In a text-generation scenario, people will often declare themselves as the author, and not the AI who generated the text for them. This is the case even when they feel that the ownership lies with the AI. %

Study 1 gave a first indication of the \aighostwriter{} in a postcard-writing scenario: For all \ims{} where participants received AI support, they tended to state that the text should belong to the AI (H1.1). Despite this, less than one-fifth of the participants declared the AI as an author of the postcard (H1.2). 
The study also showed that the sense of control and leadership were positively correlated with the level of influence over the text generation, which ranged from receiving a non-editable AI text (\getting{}) to writing the text without AI support (\writing{}; H2).
It did not make a difference for the sense of ownership and the declaration of authorship whether a participant received a personalized or placebo-personalized generated text (H3). Therefore, it was not necessary to include \textsc{Personalization} as an independent variable in Study 2.

With the pre-registered Study 2, we were able to replicate the \aighostwriter{} in a larger sample and to compare it to human ghostwriting.  In a Wizard-of-Oz design, the same placebo-personalized texts were presented as either being produced by an ``AI''or by a ghostwriter introduced as a person named ``Sasha''.
The participants' sense of ownership was significantly lower when they received a supposedly human-written text than an AI-written text. Together with the participant statements, this suggests that the AI is seen as a tool rather than an independent co-author (H4).
In line with the tendency of humans to more readily exploit AI than a human \cite{karpus_algorithm_2021}, the human ghostwriter was also credited more often than the AI ghostwriter (H5).
Study 2 provided participants with a multiple-choice list of possible options to choose from, rather than an open author declaration. Thus, they could immediately see that it was possible to choose the AI as an author. Nonetheless, there were still cases where they did not acknowledge the AI or human ghostwriter (H6).

Our findings extend the research on human-AI collaboration in writing. Specifically, we focus on cognitive and behavioral implications and on interaction design.
They also motivate recommendations for the design of AI-text-generation interfaces and for the extension of authorship frameworks to support text production.

\subsection{Implications for Cognition and Behavior in Writing}

When AI systems are introduced to the writing process, they can impact cognitive and behavioral patterns in text production, such as changes in control and the dynamics of the established writing process that, in turn, change ownership and authorship for users.

\paragraph{Involvement and Psychological Ownership}
With automation introduced by interactive AI, influence over the interaction is one core issue \cite{bergstrom_sense_2022,shneiderman_bridging_2020}. Existing research, such as \citet{lehmann_suggestion_2022}, found a change in perceived authorship depending on the control over texts, for example, when composing text with AI-generated phrases and steering automatic AI writing. We extend this work by examining authorship and ownership in more detail, personalizing the text generation, and abstracting beyond the sentence-level interaction.

Be reminded that levels of influence did not affect the sense of ownership directly in Study 1 but that the sense of ownership did change as a function of the sense of control over the content and sense of leadership in the interaction which was affected by our \ims{}. The regression models, taking into account individual variation with respect to control, highlight that there is a large variation in the mental model of human-AI interaction with respect to user involvement. This is in line with \citet{pierce_state_2003} who found that ownership increases with one's own subjective involvement. Thus, not only objective control, e.g., as induced by different \ims{}, but the subjective experience of control in human-AI interaction is essential to understanding how users relate to AI-generated content in terms of ownership. 

\paragraph{AI Support and the Cognition of Writing}
Including text suggestion tools in writing tasks also shapes the cognitive processes involved in writing. Originally, the commonly used Flower and Hayes model \cite{flower_cognitive_1981} describes writing as an iterative process comprising the subprocesses planning, translating, and revising at different levels of granularity, all internal to the writer. Hayes' revised model \cite{Hayes2012} extends and adapts this model, e.g., by introducing the ``Task Environment'' as an element. This also acknowledges the role of technology in transcription, but its position is still that of a tool at the writer's command. \citet{Bhat2022} argue that, by now, suggestion systems surpass this passive role. In fact, they found that suggestion systems influence what people write, even when suggestions do not reflect their own opinions (see also \cite{Jakesch2023opinionatedLLMs}). Consequently, they propose an adaptation of Hayes' model with interconnections between the system and the text written so far. The active role of AI systems also has consequences for authorship: if the text is influenced by a system, can an author still take full responsibility?

\paragraph{Algorithm Exploitation}
We found a large gap between the declaration of authorship in human-AI interaction when compared to human-human interaction (Study 2). Relative to a human writer, our participants mentioned AI support less often. This is in line with recent studies of algorithm exploitation. \citet{karpus_algorithm_2021} have found that while users of automated vehicles are well aware of traffic rules and the role of automated vehicles on the road, drivers are keen on exploiting automated cars as compared to cars driven by human drivers.
Thus, we find that algorithm exploitation also holds for text production with AI concerning ghost-authorship. Nevertheless, how algorithm exploitation develops and how it can be mitigated still needs to be researched more closely.

\subsection{Designing With and For Perceived Authorship}
The \aighostwriter{} has implications for interaction design:
Human-centered design for AI writing tools should consider (perceived) authorship as a design dimension. A fundamental next step is to gain more insight into the relevant UI characteristics. These might include levels of control \cite{shneiderman_human-centered_2020}, roles in the writing process~\cite{lehmann_suggestion_2022}, supporting creators \cite{Gero2019metaphor}, features for reflecting on produced text, representation of AI (e.g. anthropomorphized or not, considering the results for AI versus human ghostwriter in Study 2), and how users can attribute authorship after finishing writing (see \autoref{sec:imp_for_auth_dec}).

For instance, as our UI comparison in Study 1 suggests, the UI would at least need to afford manual edits of AI-written text to facilitate perceived authorship and control on the user's part, if that is a design goal in a particular use case. While \getting{} put the model in full control over writing the text, providing editing options shifted perceived control and leadership towards the user, and descriptively also perceived ownership (slightly). Beyond our UIs here, displaying more than one suggestion might also affect this perception if \textit{combined} with editing, as users could then express more ``editorial decisions'' overall. More generally, rich GUI components could be studied that allow users to make decisions \textit{about} the AI in advance or intermittently: For example, a GUI could offer item pickers to adjust the sentiment, scope, or length of the text (cf. the UI of \textit{Grammarly}\footnote{\url{https://app.grammarly.com/}}). A future study could examine how such controls over the generative process (vs. the final result) affect the effects observed here.

\subsection{Implications for UI Design for Authorship Declaration}
\label{sec:imp_for_auth_dec}

Besides implications on UI design for writing with AI, our findings also motivate a closer look at UI design for attributing authorship. In our studies, we used a free text field (Study 1) and a drop-down list (Study 2) -- but the design space is much larger: For instance, let us assume that a design goal is \textit{transparency} about the use of AI text generation. In this case, we might take inspiration from (early) mobile email ``disclaimers'' (e.g. ``Written on the go, may contain typos'' or just ``Sent from my iPhone''). Similarly, authors might end with an annotation such as ``Written with model X'', which would reveal AI influence on the level of the whole text. 

On a deeper level, such as sentences, interactions following the ``details-on-demand'' concept from information visualization~\cite{Shneiderman2003eyes} might reveal more: For instance, news articles typically list author names. They could be extended such that clicking on a name highlights the parts of the texts authored by that person. ``AI generated'' might then be simply another option, besides human names. Note that this does not necessarily imply that the AI is presented as on par with human authors. Therefore, declaring AI in the writing process should be designed on the user's mental model.

\subsection{Extending Contribution Frameworks for Human-AI Collaboration}

We found that when authors are given the explicit option to add AI (Study 2), then they are more likely to do so than when they are not (Study 1). Echoing this, researchers in recent publications struggle with how they should report text generation \cite{kung_performance_2022} and policies vary from banning text generation to mandatory labeling of AI-based text generation \cite{10.1001/jama.2023.1344}. Our data cannot support this simple approach to the problem. For example, we found that perceived ownership heavily varies with subjective controllability when using the AI model. Thus, a more continuous approach to AI declaration may be needed to fit the writer's mental model. The CRediT taxonomy \cite{allen2014publishing,brand2015beyond,allen2019can} for roles in authorship contribution serves the very purpose of finely declaring author involvement in different stages of scientific manuscript production (e.g., A wrote the draft, B supervised the study). Adapting frameworks like CRediT to suit the contributions of algorithms could enrich the debate from a human-centered perspective grounded in common practices. 

In a broader perspective, the perceived importance of such authorship declarations may vary depending on culture and (emerging) social norms and expectations around the use of AI. Investigating these aspects further in the future could link our work here, for example, with research directions on AI-mediated communication and related social and cultural dimensions of such technology (e.g., \cite{Goldenthal2021CiHB, Hancock2020JCMC, Mieczkowski2021cscw}).

For example, in paper-based personal communication, the handwriting usually marks authorship---even if only as a handwritten note added to a pre-printed greeting card.
The emergence of services such as postcard apps, where senders enter the text that is then printed and sent (e.g., MyPostcard\footnote{\url{https://www.mypostcard.com/en/}}), already means that receivers cannot fully trace authorship. We expect senders to increasingly resort to AI support for generating such texts.
These changing practices will initiate discussions on the declaration and validation of authorship in personal communication contexts.

\subsection{Limitations}
\label{sec:limitations}

Some limitations have to be taken into account with regard to the scenario, the personalization, and the interfaces used in the studies. 
First, the scenario of writing a postcard could be judged as being somewhat artificial and simple, and ideally, future studies should replicate the \aighostwriter{} in other writing tasks that require longer and more nuanced texts. However, previous studies have often used academic writing tasks (e.g., \cite{nylenna_authorship_2014,allen_how_2019}), where personalized writing might not be as relevant, and where the context may not generalize with regard to authorship -- given that in academic settings, there are practices and rules in place.
In contrast to writing postcards in real life, the postcards were not sent to someone but published online. They also did not describe an actual trip the participants had made. This may have influenced how the participants decided on declaring authorship. Nonetheless, we assume that not sending the postcards actually increases the probability of declaring AI as an author because personal stakes are lower.

Second, we did only personalize the generated texts for half of the participants in Study 1 and not at all for Study 2. Moreover, the number of prompt-completion pairs used for fine-tuning was lower than what is recommended by OpenAI\footnote{\url{https://platform.openai.com/docs/guides/fine-tuning}}. One could thus assert that true personalization might produce different results. However, we found no differences between true and placebo-personalization in terms of the sense of ownership and declared authorship.
This suggests that the \aighostwriter{} does not depend on the quality of personalization but only on whether participants can relate to the text generated. Likewise, some participants were not fully satisfied with the level of personalization. Nevertheless, with the increase in the quality of personalization in AI-generated texts, the size of the \aighostwriter{} should further increase. Thus, we deem that the choice of placebo-personalization does not limit our results but rather specifies that the existence of the effect is not tied to the quality of personalization. 

Third, we found strong differences in author declaration between both studies. In Study 1, only about 20\% of participants mentioned AI in the free text field underneath the postcard. In Study 2, about 70\% mentioned AI when choosing from a selection menu. Future studies may closely investigate how to design interfaces of authorship declaration so that contributions of AI systems are declared. Here, future studies could also look at whether GPT or other LLMs might be disclosed as tools used while writing, e.g., in the methods section of a scientific article. 

Fourth, our studies can merely describe how people \textit{do declare} authorship for personalized AI-generated texts. As interdisciplinary efforts of cognitive science and HCI may develop a prescriptive framework that can recommend how people \textit{should declare} authorship when supported by AI, it remains open how people will apply these prescriptive frameworks. 

Fifth, for the sake of brevity, the data analyzed in this study is purely quantitative and can thus only describe single points of judgment, impressions, and behavior. A more qualitative approach to studying writing, following \citet{Bhat2022}, could paint a more nuanced picture with regard to the reasoning of participants in crediting or not crediting AI in the writing of a personalized text. 

Sixth, we focused on (placebo-)personalized AI-generated texts as they mirror the process of ghostwriting. Therefore, our results may be biased in favor of attributing authorship to oneself. Whether the discrepancy also holds for interaction in non-personalized AI writing contexts still needs to be investigated.

\section{Conclusion and Outlook}

Our findings based on two empirical studies with a total of 126 participants reveal the \aighostwriter{}; we show that people often do not disclose AI as an author when using personalized AI-generated texts in writing, although they attribute ownership to the AI. This effect was independent of interaction methods and reduced when switching from open-ended author declaration fields to predefined response suggestions. Comparing human and AI ghostwriters, we found attributing authorship to oneself to be more prevalent when writing with AI support. 
Given the differences in perceived ownership and author declaration, we call for the development of authorship attribution frameworks that take the user and their relation to the generative model into account.

\bibliographystyle{ACM-Reference-Format}
\bibliography{ghostwriter_ai}

\begin{acks}

We thank Otso Haavisto for providing comments on an earlier draft. This project is funded by the Bavarian State Ministry of Science and the Arts with the Bavarian Research Institute for Digital Transformation (bidt) and also funded by the HumanE AI Network from the European Union’s Horizon 2020 research and innovation program under grant agreement No. 952026.

\end{acks}

\appendix

\section{Study 1: Full List of Measures}
\label{sec:measures_study1}

\begin{longtable}{p{\textwidth}}
    \caption{Measures applied in Study 1, Parts 2 and 3. The fine-tuning questions asked in Part 1 are provided as supplementary material on OSF: \url{https://osf.io/n4svx/?view_only=916873e81d244be6be8e0790531b1197}. Unless otherwise mentioned, we used 7-point Likert scales (1 – strongly disagree, 7 - strongly agree). When necessary, questions were slightly modified to be applicable in the context of text generation (e.g. ``used'' instead of ``worn'' for the creepiness scale).}
    \label{tab:study1_measures} \\

        \toprule
        \multicolumn{1}{c} {\textbf{Part 2a: Before the system interaction}} \\
        \midrule
        \endfirsthead
        \endhead

        \textbf{Prior experience}
        \begin{itemize}
            \item Please list all previous experiences you have in writing with generated text \newline
            (none – Writing with word or sentence suggestions – Writing with auto-completion - Writing with auto-correction - Using the smart reply feature - Other (+ text field))
            \item Are texts sometimes written on your behalf by other people, for example by an assistant at work who prepares texts, presentations, or speeches for you? (Yes / No)
        \end{itemize} \\
        \midrule

        \textbf{Mood}
        \begin{itemize}
            \item How is your mood today? (Rating: sad smiley 1 -- happy smiley 5)
        \end{itemize} \\

        \toprule
        \multicolumn{1}{c} {\textbf{Part 2b: Repeated for each interaction method (IM)}} \\
        \midrule
        
        \textbf{Sense of Ownership}
            \begin{itemize}
                \item Questions on perceived authorship and ownership derived from the authorship criteria defined by the International Committee of Medial Journal Editors\footnote{\url{https://www.icmje.org/recommendations/browse/roles-and-responsibilities/defining-the-role-of-authors-and-contributors.html}} and \citet{nylenna_authorship_2014}:
                \begin{itemize}
                    \item I am the main contributor to the content of this postcard. 
                    \item I have made substantial contributions to the content of the text.
                    \item I drafted this postcard.
                    \item I revised this postcard critically.
                    \item I have given final approval of this text being uploaded.
                    \item I am accountable for all aspects of the text.
                    \item I am responsible for at least part of the text.
                    \item My name should appear underneath this postcard.
                \end{itemize}
                \item I feel like I am the author of the text. \cite{lehmann_suggestion_2022}
                \item To whom should this text belong? (Slider: Me -- AI)
            \end{itemize} \\
        \midrule
       
        \textbf{Sense of Leadership}
        \begin{itemize}
            \item I felt like I was writing the text and the artificial intelligence was assisting me. \cite{lehmann_suggestion_2022}
            \item I felt like the artificial intelligence was writing the text and I was assisting. \cite{lehmann_suggestion_2022}
            \item Who took the lead in writing during this interaction method? (Slider: Me -- AI)
        \end{itemize} \\
        \midrule
        
        \textbf{Sense of Control}
        \begin{itemize}
            \item Who was in control of the content? (Slider: Me -- AI)
            \item Sense of Agency Scale \cite{tapal_sense_2017}
                \begin{itemize}
                    \item I was just an instrument in the hands of the AI system. 
                    \item The resulting text just happened without my intention.
                    \item I was the author of my actions while interacting with the system.
                    \item The consequences of my actions felt like they don’t logically follow my actions while interacting with the system. 
                    \item The outcomes of my actions generally surprised me. 
                    \item While I am interacting with the system, I feel like I am a remote controlled robot.
                    \item I am completely responsible for the resulting text.
                \end{itemize}
            \item It felt like I was in control of the text during the task. \cite{bergstrom_sense_2022}
            \item I felt like the AI system was acting as a ghostwriter, writing the postcard on my behalf.
            \item I felt like the AI system was acting as a tool which I could control.
        \end{itemize} \\
        \midrule
        
        \textbf{Text match}
            \begin{itemize}
                \item I would actually send this postcard to my friends and family.
                \item I would have written a similar postcard by myself.
                \item The postcard mostly contains words and/or phrases that I usually use when writing in English. \cite{lehmann_suggestion_2022}
                \item I am satisfied with the text. \cite{lehmann_suggestion_2022}
            \end{itemize} \\ %
        \midrule
        
        \textbf{User Experience}
            \begin{itemize}
                \item AttrakDiff adjective pairings on a 7-point Likert scale as used on \url{https://www.attrakdiff.de/index-en.html}, see also \cite{hassenzahl_attrakdiff_2003}: %
                    \begin{itemize} %
                        \item \textit{pragmatic quality}: human -- technical, simple -- complicated, practical -- impractical, cumbersome -- straightforward, predictable -- unpredictable, confusing -- clearly structured, unruly -- manageable
                        \item \textit{hedonic quality -- stimulation}: inventive -- conventional, unimaginative -- creative, bold -- cautious, innovative -- conservative, dull -- captivating, undemanding -- challenging, novel -- ordinary
                        \item \textit{hedonic quality -- identity}: connective -- isolating, unprofessional -- professional, tacky -- stylish, premium -- cheap, integrating -- alienating, separates me from people -- brings me closer to people, presentable -- unpresentable
                        \item \textit{hedonic quality -- appeal}: pleasant -- unpleasant, ugly -- attractive, likeable -- disagreeable, rejecting -- inviting, good -- bad, repelling -- appealing, motivating -- discouraging
                    \end{itemize}
                \item How much did you like this postcard? (Rating: 1-5 stars)
                \item How hard was it to understand and use this method? (Very difficult - very easy)
                \item I had fun while interacting with the system.
                \item Open questions on the interaction method
                    \begin{itemize}
                        \item How could this interaction method be improved?
                        \item What are the advantages of this method with respect to the writing interaction?
                        \item What are the disadvantages of this method with respect to the writing interaction?
                    \end{itemize}
                \item Specific questions for individual interaction methods
                    \begin{itemize}
                        \item \choosing{}: Would you have liked to have more than three texts to choose from?
                        \item \choosing{}: What would be a good number of texts to choose from?
                        \item \choosing{}: Why did you choose your selected postcard and not one of the two other postcards?
                        \item \choosing{}: Which strategies did you apply when selecting one of the postcards?
                        \item \editing{}: Did you want to change more words than which was possible in this task?
                        \item \editing{}: Number of words changed
                        \item \editing{}: Change logs (e.g., added, deleted, or changed words or punctuation)
                    \end{itemize}
            \end{itemize} \\

        \toprule
        \multicolumn{1}{c} {\textbf{Part 2c: Overall assessment after all IMs}} \\
        \midrule        
        
        \textbf{Mood}
        \begin{itemize}
            \item The participants' current mood \cite{schone_experiences_2019}
        \end{itemize} \\ %
        \midrule
        
        \textbf{Creepiness}
            \begin{itemize}
                \item The Perceived Creepiness of Technology Scale (PCTS) for the system in general \cite{wozniak_creepy_2021}
            \end{itemize} \\
        \toprule
        \multicolumn{1}{c} {\textbf{Part 3: Two days after the interaction}} \\
        \midrule
        
        \textbf{Memory}
            \begin{itemize}
                \item Sleep and mood to identify possible impacts on the participants' memory \cite{schone_experiences_2019,rasch_about_2013}
                \item Text remembrance based on \cite{kisker_virtual_2021}
                     \begin{itemize}
                         \item What do you recognize this text as? (Vividly remembered - Familiar - Rather unknown - Definitely unknown)
                         \item How was this text created? (Text was written by myself - Text was generated by AI and edited by me - Text was generated by AI and chosen by me - Text was generated by AI - I don't know)
                     \end{itemize}
            \end{itemize} \\
        \midrule
        \textbf{Retrospective assessment of the postcard quality}
            \begin{itemize}
                \item How much did you like this postcard? (Rating: 1-5 stars)
            \end{itemize} \\
        \midrule
        \textbf{The role of AI in writing} \\
            Open questions on the interaction methods and the role of AI technology in general
            \begin{itemize}
                \item Advantages of automatic text generation by an AI
                \item Disadvantages/risks of automatic text generation by an AI
                \item Use such technologies in the future
                \item Use cases for AI to generate text
                \item Should it be mandatory to mark texts that were created with the help of AI as such (Yes - No - Other) + open text field for reasons
                \item Feedback
            \end{itemize}
        \\
        \bottomrule
\end{longtable}

\section{Study 2: Full List of Measures}

\begin{longtable}{p{\textwidth}}
    \caption{Measures applied in Study 2. Unless otherwise mentioned, we used 7-point Likert scales. (1 – strongly disagree, 7 - strongly agree)}
    \label{tab:study2_measures} \\
        \toprule
        \multicolumn{1}{c} {\textbf{Part 1}} \\
        \midrule
        
        \textbf{Prior experience}
            \begin{itemize}
                \item Please list all previous experiences you have in writing with generated text \newline
            (none – Writing with word or sentence suggestions – Writing with auto-completion - Writing with auto-correction - Using the smart reply feature - Other (+ text field))
                \item Are texts sometimes written on your behalf by other people, for example by an assistant at work who prepares texts, presentations, or speeches for you? (Yes / No)
            \end{itemize} \\
        \midrule
        
        \textbf{Attitudes towards AI}
            \begin{itemize}
                \item Perceived relevance and usefulness of AI \cite{hong_ai_2021}
                    \begin{itemize}
                        \item AI is a positive force in the world.
                        \item AI research should be funded more.
                        \item AI is generally helpful.
                        \item There is a need to use AI.
                    \end{itemize}
                \item Perceived AI competence \cite{hong_ai_2021}. How would you rate your confidence in the following:
                    \begin{itemize}
                        \item Explaining what artificial intelligence is.
                        \item Having a conversation about artificial intelligence.
                        \item My knowledge about artificial intelligence.
                    \end{itemize}
                \item Creativity of AI as assessed in \cite{hong_are_2021}
                    \begin{itemize}
                        \item I think AI can be creative on its own.
                        \item I believe AI can make something new by itself.
                        \item Products developed by AI can be considered as creative works.
                    \end{itemize}
            \end{itemize} \\

        \midrule
        \multicolumn{1}{c} {\textbf{Part 2a: One iteration each for the human- and AI-generated text (counterbalanced)}} \\
        \midrule
        \textbf{Sense of Ownership} \\
            As in Study 1 \\
        \midrule
       
        \textbf{Sense of Leadership} \\
            As in Study 1 \\
        \midrule
        
        \textbf{Sense of Control} \\
            As in Study 1 \\
        \midrule

        \textbf{Text match}
            \begin{itemize}
                \item Questions from Study 1
                \item I felt like the text was written for me personally.
            \end{itemize} \\
        \midrule
        
        \multicolumn{1}{c} {\textbf{Part 2b: After both texts}} \\
        \midrule
        \textbf{Attitudes towards AI}
            \begin{itemize}
                \item Questions on relevance, usefulness, creativity of AI and participants' AI competence as in Study 2, Part 1 
                \item Godspeed questionnaires on \textit{anthropomorphism} and \textit{intelligence} \cite{bartneck_measuring_2017}, as used in \cite{lermann_henestrosa_automated_2023}
            \end{itemize} \\

        \midrule
        \textbf{Open questions on authorship}
        \begin{itemize}
            \item How did you decide what author attribution should be added to the postcard written by <AI/Sasha>?
            \item In \textbf{your opinion}, should it be mandatory to mark texts that were created with the help of an <AI/human> as such?
            \item Do you think that \textbf{other people} expect you to attribute <AI/Sasha> in your text? Why?
        \end{itemize} \\
        \midrule
        \textbf{Preference}
            \begin{itemize}
                \item Which postcard did you like better? (The one written by the AI - The one written by Sasha - I'm indifferent)
            \end{itemize} \\
        \bottomrule

\end{longtable}

\section{Study 1: Blog-Style Website}

\begin{figure}[hb]
    \centering
    \includegraphics[width=\textwidth]{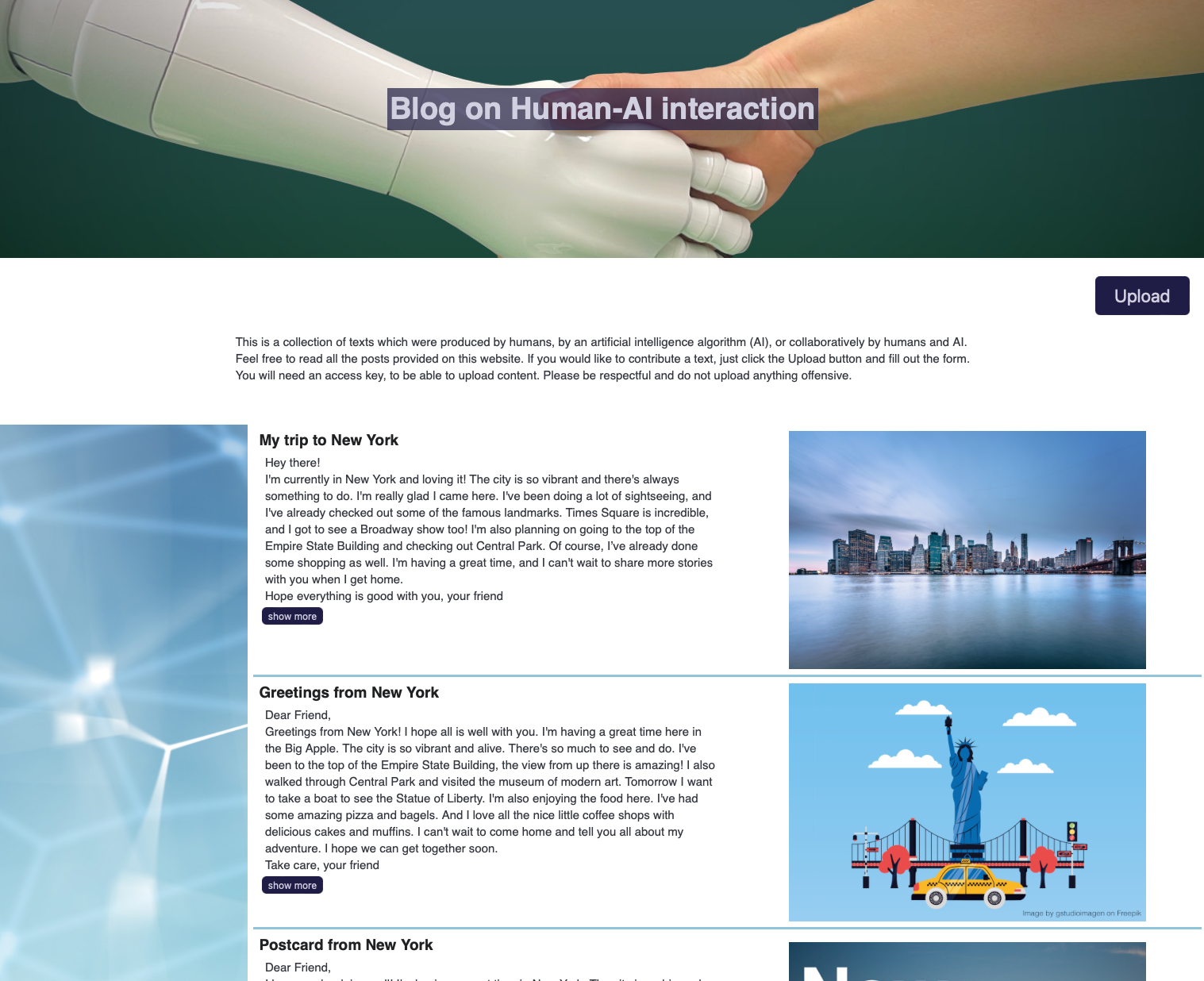}
    \caption{Screenshot of the blog-style website we created. Image credits: network background by kjpargeter on Freepik. Top postcard by Chris Schippers on Pexels. Bottom postcard by gstudioimagen on Freepik. Header image authors' own.}
    \label{fig:screenshot_blog}
\end{figure}

\begin{figure}[bh]
    \centering
    \includegraphics[width=\textwidth]{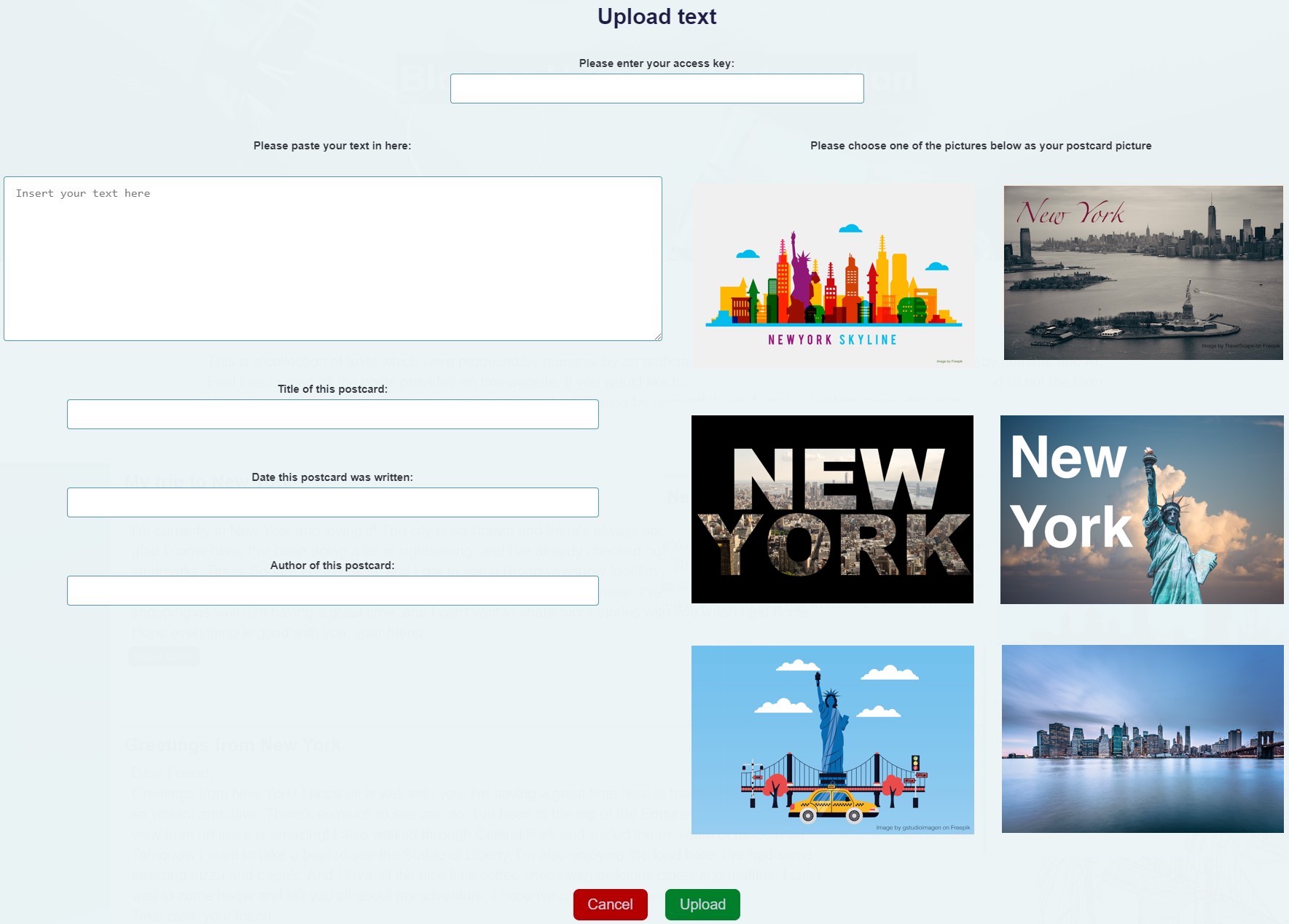}
    \caption{Screenshot of the upload form on the blog-style website we created for Study 1. Original image credits: Top left by Freepik. Top right by TravelScape on Freepik. Center left by Pixabay (photograph below text). Center right by ParentRap on Pixabay. Bottom left by gstudioimagen on Freepik. Bottom right by Chris Schippers on Pexels.}
    \label{fig:screenshot_upload}
\end{figure}

\end{document}